\documentclass[a4paper,11pt,nobalancelastpage,superscriptaddress]{revtex4-2}
\UseRawInputEncoding
\usepackage{color,ifthen,amsthm,amsmath,amsxtra,amsfonts,dsfont,graphicx,bm,tikz,scalerel,wasysym,bbm,graphicx,amsthm,braket,physics,empheq}
\usepackage[colorlinks=true,linkcolor=blue, citecolor=blue, urlcolor=blue, bookmarks]{hyperref}

\usepackage[T1]{fontenc}
\usepackage[utf8]{inputenc}

\newcommand{\be}{\begin{equation}}
\newcommand{\ee}{\end{equation}}
\newcommand{\bea}{\begin{eqnarray}}
\newcommand{\eea}{\end{eqnarray}}
\newcommand{\bse}{\begin{subequations}}
\newcommand{\ese}{\end{subequations}}

\theoremstyle{plain}

\theoremstyle{plain}

\theoremstyle{plain}

\begin{document}

\title{Dynamical symmetry restoration in the Heisenberg spin chain}

\author{Colin Rylands}
\affiliation{SISSA and INFN Sezione di Trieste, via Bonomea 265, 34136 Trieste, Italy}
\author{Eric Vernier}
\affiliation{Laboratoire de Probabilités, Statistique et Modélisation CNRS, Université Paris Cité, Sorbonne Université Paris, France}

\author{Pasquale Calabrese}
\affiliation{SISSA and INFN Sezione di Trieste, via Bonomea 265, 34136 Trieste, Italy}
\affiliation{International Centre for Theoretical Physics (ICTP), Strada Costiera 11, 34151 Trieste, Italy.}

\begin{abstract}

The entanglement asymmetry is an observable independent tool to investigate the relaxation of quantum many body systems through the restoration of an initially broken symmetry of the dynamics.  In this paper we use this to investigate the effects of interactions on quantum relaxation in a paradigmatic  integrable model.  Specifically, 
we study the dynamical restoration of the $U(1)$ symmetry corresponding to rotations about the $z$-axis in the XXZ model quenched from a tilted ferromagnetic state.  We find two distinct patterns of behaviour depending upon the interaction regime of the model.  In the gapless regime, at roots of unity, we find that the symmetry restoration is predominantly carried out by bound states of spinons of maximal length.  The velocity of these bound states is suppressed as the anisotropy is decreased towards the  isotropic point leading to slower symmetry restoration.  By varying the initial tilt angle,  one sees that symmetry restoration is slower for an initally smaller tilt angle, signifying the presence of the quantum Mpemba effect.  In the gapped regime however,  spin transport for non maximally tilted states, is dominated by smaller bound states with longer bound states becoming frozen.  This leads to a much longer time scales for restoration compared to the gapless regime.  In addition, the quantum Mpemba effect is absent in the gapped regime. 

\end{abstract}

\maketitle

\section{Introduction}
\label{sec:introduction}
A closed,  far-from-equilibrium,  quantum system will generically relax locally to a stationary state, leading to the emergence of statistical mechanics. This fact has been established since the turn of the millennium and verified in numerous theoretical and experimental settings~\cite{srednicki1994chaos,polkovnikov2011nonequilibirum,
rigol2008thermalization,calabrese2016introduction, VidmarRigol, essler2016quench, doyon2020lecture, bastianello2022introduction, alba2021generalized,rylands2020nonequilibrium}. The route the system takes to relaxation is much trickier to pin down.  Traditionally, one might study this through the expectation values of  local observables.  Such quantities are, however, notoriously difficult to calculate in far from equilibrium systems, even in integrable models~\cite{korepin1997quantum, caux2016quench}.  What's more, the results are necessarily observable dependent, making it difficult to discern universal properties of quantum relaxation processes.  To overcome these issues and obtain universal information, an observable independent, symmetry based quantifier of local relaxation has been developed. This quantity is called the entanglement asymmetry~\cite{ares2022entanglement}. Using the methods of entanglement entropy,  it tracks the restoration of symmetry, at the subsystem level,  which is broken in the initial state.  So far this approach has been quite useful in uncovering a number of new and unexpected phenomena such as the absence of  $U(1)$ symmetry  restoration in quenches of spin chains from certain initial states~\cite{ares2023lack} or the emergence of effective subsystem conservation laws in Haar random states~\cite{ares2023entanglement}.  Perhaps the most prominent feature discovered using the entanglement asymmetry is the quantum Mpemba effect~\cite{ares2022entanglement}.  It is a quantum analogue of the classical effect wherein a non-equilibrium state can relax faster if it is initially further from equilibrium~\cite{mpemba1969cool,ahn2016experimental,hu2018confirmation, lasanta2017mpemba, klich2019mpemba,kumar2020exponentially,kumar2022anomalous, walker2023Mpemba, teza2023relaxation, walker2023optimal, bera2023effect}.  The quantum effect has now been studied in numerous scenarios and, although definitively a non generic behaviour,  is surprisingly ubiquitous and robust, appearing in integrable~\cite{murciano2024entanglement,ferro2023nonequilibrium,rylands2024microscopic,bertini2024dynamics,chalas2024multiple,
ares2024quantum,yamashika2024entanglement}, and chaotic dynamics~\cite{liu2024symmetry,turkeshi2024quantum,caceffo2024entangled,klobas2024translation,liu2024quantum,foligno2024nonequilibrium} as well as experimentally in a trapped ion quantum simulator~\cite{joshi2024observing} (See also~\cite{nava2019lindblad,nava2024mpemba,kochsiek2022accelerating,carollo2021exponentially,manikandan2021equidistant,ivander2023hyper,chatterjee2023quantum,chatterjee2024multiple,strachan2024nonmarkovian,moroder2024thermodynamics,aharony2024inverse,zhang2024observation} on distinct but related Mpemba effects in quantum systems).  The entanglement asymmetry, or simply asymmetry for short,  has also been utilized in studying symmetry breaking in equilibrium many body systems~\cite{capizzi2023entanglement,capizzi2023universal,khor2023confinement,chen2024renyi,fossati2024entanglement,lastres2024entanglement,benini2024entanglement}.

In this paper we perform a systematic study of the effects of interactions on the dynamical restoration of a $U(1)$ symmetry following a quantum quench.  Specifically,  we study dynamics of the entanglement asymmetry in the XXZ spin chain quenched from a tilted ferromagnetic state which breaks the $U(1)$ symmetry corresponding to the magnetization along the $z$ direction.  As previously reported in~\cite{ares2023entanglement},  at the XX point the symmetry is restored at long times while the quantum Mpemba effect is observed when comparing different initial tilt angles -- the larger the initial tilt angle, the more the symmetry is broken and the quicker it is restored.  Upon introducing interactions we observe a distinct departure in this behaviour depending on whether the model is in the gapless or gapped regime.   
In the gapless regime, as one moves away from the free point we see that the symmetry is also restored and the quantum Mpemba effect occurs but the time taken to achieve this increases.  As the isotropic point is approached, the time scale for symmetry restoration diverges and the Mpemba effect is pushed to later and later times.  Exactly at the isotropic point the tilted ferromagnet becomes an eigenstate of the model and no symmetry restoration or in fact dynamics takes place.  Within the gapped regime, the behaviour is markedly different.  Symmetry restoration takes an exceedingly long time for non maximal tilt angles, while for a fully tilted ferromagnet no restoration takes place on ballistic scales. Notably, the quantum Mpemba effect disappears in the gapped regime -- the more symmetry is broken the longer it takes to be restored.

This behaviour can be understood using the quasiparticle picture as well as the dressing of quasiparticle properties in interacting integrable systems.  The elementary excitations of the XXZ chain are spinons and multi-spinon bound states called strings~\cite{takahashi1999thermodynamics},   the number of spinons bound together being referred to as the string length.  At special points in the gapless regime when the anisotropy parameter is a root of unity, the maximal length of the string is restricted to be a finite value but which increases as the isotropic point is reached.  In the gapped regime, however, there is no such restriction on the maximal string length.  The length of the string coincides with the bare magnetization that a string carries, however this is dressed by the interactions and depends on the particular state of the system.  For quenches from the tilted ferromagnet in the gapless regime,  the magnetization is predominantly carried by the longest strings which necessarily have a slower velocity.  The magnetization of shorter strings is significantly dressed towards zero.  Indeed,  for the fully tilted ferromagnet the dressed magnetization of all but the last two strings vanishes exactly.  Accordingly,  as the isotropic point is approached,  the restoration of the symmetry is shifted to later and later times since magnetization is concentrated on longer and longer strings.  As we enter the gapped regime there is no longer a restriction on the maximal string length but the magnetization is still primarily determined by the longest strings. The symmetry restoration, therefore relies upon the transport of very large strings but whose velocity is drastically suppressed by dressing.  In the fully tilted case, the dressed magnetization of all strings is vanishing and there is no ballistic transport of spin~\cite{znidaric2011spin,karrasch2017real,illievski2018superdiffusion,gopalakrishnan2019kinetic,bertini2021finite}.  Thus the symmetry cannot be restored on ballistic times scales.   This lack of symmetry restoration on the ballistic scale for the fully tilted state and the exceedingly long timescale for restoration at smaller tilt angles then conspire to destroy the quantum Mpemba effect.   Microscopically,  the spin Drude weight in the gapless regime is dominated by slower and slower quasparticles as the tilt angle is decreased. This results in slower spin transport and symmetry restoration. In the gapped case however, there is no such shift toward slower quasiparticles.  Consequently, no quantum Mpemba effect takes place. 

The remainder of the paper is structured as follows.  In section~\ref{sec:model} we introduce the model, initial state and review some of its basic properties.  In section~\ref{sec:entangelementasymmetry} we discuss our main quantity of interest, the entanglement asymmetry.  We define it in a general way before specializing to the case at hand and discuss how it can be calculated in integrable models.  In section~\ref{sec:gapless} we present results for the quench in the gapless regime, paying particular attention to the fully tilted initial state.  In section~\ref{sec:gapped} we present results for the quench in the gapped regime and show the absence of the quantum Mpemba effect in the ballistic scaling regime.  In section~\ref{sec:conslusions} we summarize and conclude. 

\section{Model and Quench}
\label{sec:model}

The XXZ spin chain has been one of the most important models for understanding the behaviour of many-body quantum systems and in particular the quench dynamics of interacting integrable models~\cite{wouters2014quenching,pozsgay2014correlations,brockmann2014quench,liu2014quench,illievski2015complete}.  Its Hamiltonian is
\begin{equation}\label{eq:Hamiltonian}
    H=\sum_{j=1}^L S^x_{j}S^x_{j+1}+ S^y_{j}S^y_{j+1}+\Delta  S^z_{j}S^z_{j+1}
\end{equation}
where $S^{x,y,z}_{j}$ are spin half operators acting on the site $j$.  $\Delta$ is known as the anisotropy, $L$ is the size of the chain and we assume periodic boundary conditions, $S^{x,y,z}_{j}=S^{x,y,z}_{j+L}$.  The model has a $U(1)$ conserved charge, namely the $z$-component of the total spin $S^z=\sum_{j=1}^L S^z_j$ which is enhanced to full $SU(2)$ symmetry at the isotropic point $\Delta=1$.  The regimes $|\Delta|\leq 1$ and $\Delta>1$, are referred to as gapless and gapped respectively.  As suggested by their names they have markedly distinct properties, including the nature of their spectra, transport properties and underlying mathematical structure, which will be discussed further below. 

\textbf{Initial state:} We study the non-equilibrium properties of the model in the most common way, through a quantum quench. The system is initially prepared in a state $\ket{\Psi(0)}$ and allowed to undergo unitary time evolution according to~\eqref{eq:Hamiltonian}, namely,  $\ket{\Psi(t)}=e^{-iHt}\ket{\Psi(0)}$.  We take as the initial state a tilted ferromagnet
\begin{eqnarray}\label{eq:tilted_ferro}
\ket{\Psi(0)}=e^{i\theta S^y}\ket{\Uparrow}
\end{eqnarray}
where $\ket{\Uparrow}$ is the ferromagnetic state with all spins pointing in the positive $z$ direction and $S^y=\sum_{j=1}^LS^y_j$ the total spin in the $y$ direction.   Our reasons for choosing this state are manifold.  It is the simplest possible state which breaks the $U(1)$ symmetry associated to rotations about the $z$ axis possessed by $H$.  Moreover,  it is an integrable initial state~\cite{piroli2017integrable} meaning that it is related to a boundary condition in space (rather than time) which preserves integrability~\cite{sklyanin1988boundary}.  Such types of initial states are known to lead to simple expressions for their overlaps with Bethe states~\cite{brockmann2014overlaps,brockmann2014gaudin,pozsgay2014overlaps,pozsgay2018overlaps,brockmann2014neel,foda2016overlaps,gombor2020boundary,gombor2021boundary,gombor2021factorized,jiang2020exact,rylands2022integrable,rylands2023solution} which allows to calculate their quasiparticle occupation functions analytically.  Aside from being integrable and symmetry breaking, the tilted ferromagnet has two other key properties dictating its choice.  First, it is one site translationally invariant and second it has non zero magnetization.  The first point is important when considering quenches in the gapless regime.  Therein, at roots of unity,  the model possess additional conserved quantities~\cite{prosen2013families,ilievski2016quasilocal} which do not commute with $S^z$. These can have non zero expectation value on states which break the symmetry but are not one site shift invariant and prevent the restoration of the $S^z$ symmetry~\citep{ares2023lack}.  Since we wish to study symmetry restoration we should avoid such states.  
The second point, on the other hand is important when studying quenches in the gapped regime. In that case it is now well established~\cite{znidaric2011spin,karrasch2017real,illievski2018superdiffusion,gopalakrishnan2019kinetic,bertini2021finite} that in the zero magnetization sector, spin is transported diffusively and not ballistically as in the gapless regime.  Our methods will only produce nontrivial results when the charges associated to the symmetry are transported ballistically and so this issue is avoided by~\eqref{eq:tilted_ferro} at least away from $\theta=\pi/2$.  

\subsection{TBA description}
Being an integrable model, the eigenstates and spectrum of the XXZ chain are known exactly~\cite{bethe1931theorie,orbach1958linear}.  Moreover, in the thermodynamic limit the system admits a description in terms of the thermodynamic Bethe ansatz which we now briefly recap~\cite{takahashi1999thermodynamics}.  The system supports a set of stable quasiparticle excitations specified by a species index $m=1,\dots,M$ and a rapidity variable $\lambda\in[-\Lambda,\Lambda]$.  The number of species, $M$, and bounds on the rapidity, $\Lambda$ depend on the regime of the model and the particular value of the anisotropy for instance, at $\Delta>1$ $M=\infty$ and $\Lambda=\pi/2$.  

In the thermodynamic limit, a stationary state of the system is described through a set of distributions for these quasiparticles, in particular $\vartheta_m(\lambda)\in [0,1]$ is the occupation function of quasiparticle of species $m$ at rapidity $\lambda$, $\rho^t_m(\lambda)$ is their density of states and $\rho_m(\lambda)=\vartheta_m(\lambda)\rho^t_m(\lambda)$ is the distribution of occupied modes.   These distributions are coupled to each other via a set of integral equations whose form also depends on the parameter regime.  
Each quasiparticle carries a bare charge $q_m$ under the $U(1)$ symmetry.  This is dressed by the interactions after which it is denoted $q_{{\rm eff}, m}(\lambda)$.  Similarly, each quasiparticle has a bare velocity which is dressed also, we denote this by $v_m(\lambda)$.   We now present details in each of the regimes, employing the following notation
\begin{eqnarray}
(f\star g)(\lambda)=\int_{-\Lambda}^\Lambda {\rm d}\mu\, f(\lambda-\mu)g(\mu). 
\end{eqnarray}

\textbf{Gapless regime:} In the gapless regime we parameterize the anisotropy via the parameter $\gamma$ defined by  $\Delta=\cos(\gamma)$ and moreover we restrict to the simplest possible case wherein $\gamma=\frac{\pi}{p+1}$.  With this choice there are $M=p+1$ quasiparticle species also known as strings, the first $p$ of which have bare magnetization or string length $q_m=m$ while the last has $q_{p+1}=1$.   
For a given stationary state defined by a set of occupation  functions $\vartheta_m(\lambda)$ the density of states $\rho^t_m(\lambda)$ and density of occupied modes $\rho_m(\lambda)$ are related by the integral equations (we drop the explicit $\lambda$ dependence to lighten the notation),
\begin{eqnarray}\label{eq:BT_1}
\rho^{t}_1&=&s + s\star\left[(1-\vartheta_2)\rho^t_{2}+\delta_{p,2} \rho_{p+1}\right]\\\label{eq:BT_2}
    \rho^t_m &=&s\star\left[(1-\vartheta_{m-1})\rho^t_{m-1}+(1-\vartheta_{m+1})\rho^t_{m+1}+\delta_{p,m+1}\rho_{p+1}\right],~~1<m<p,\\\label{eq:BT_3}
    \rho^t_p &=&\rho^t_{p+1}=s\star[ (1-\vartheta_{p-1})\rho^t_{p-1}].
\end{eqnarray} 
with $s(\lambda)=\sech{(\pi\lambda/\gamma)}/(2\gamma)$, and the bound on rapidities now being $\Lambda=+\infty$. A similar set of integral equations determine the quasiparticle velocity 
\begin{eqnarray}\label{eq:vBT_1}
v_1\rho^{t}_1&=&-\frac{\sin\gamma}{2} \partial_\lambda s
    + s\star\left[(1-\vartheta_2)v_2\rho^t_{2}+\delta_{p,2} v_{p+1}\rho_{p+1}\right]\\\label{eq:vBT_2}
    v_m \rho^t_m&=&s\star\left[(1-\vartheta_{m-1})v_{m-1}\rho^t_{m-1}+(1-\vartheta_{m+1})v_{m+1}\rho_{m+1}^t+\delta_{p,m+1}v_{p+1}\rho_{p+1}\right],~~1<m<p,\\\label{eq:vBT_3}
  v_p   \rho^t_p&=&\rho^t_{p+1}v_{p+1}=s\star[ (1-\vartheta_{p-1})v_{p-1}\rho^t_{p-1}].
\end{eqnarray}
 Note that as a consequence of~\eqref{eq:BT_3} and~\eqref{eq:vBT_3} we have that $v_p(\lambda)=v_{p+1}(\lambda)$.  The dressed magnetization $q_{{\rm eff},m}$ satisfies
\begin{eqnarray}\label{eq:qBT_1}
q_{{\rm eff},1}&=& s\star\left[(1-\vartheta_2)q_{{\rm eff},2}-\delta_{p,2}\vartheta_{p+1}q_{\rm eff,p+1}\right]\\\label{eq:qBT_2}
q_{{\rm eff},m}&=&s\star\left[(1-\vartheta_{m-1})q_{{\rm eff},j-1}+(1-\vartheta_{m+1})q_{{\rm eff},m+1}-\delta_{p,m+1}\vartheta_{p+1}q_{{\rm eff},p+1}\right],~~1<m<p,\\\label{eq:qBT_3}
    q_{{\rm eff},p} &=&p+1-q_{{\rm eff},p+1} =\frac{p+1}{2}-s\star[ (1-\vartheta_{p-1})q_{{\rm eff},p-1}].
\end{eqnarray} 
The occupation functions are required as inputs for all these equation. To find them 
we make use of the integrability of the tilted ferromagnetic state which allows us to determine $\vartheta_m(\lambda)$ analytically.   The details are quite involved and are presented in full in the appendix~\ref{sec:appendix}.  As an example, for $p=1$, corresponding to the XX model, we have
\begin{eqnarray}
\vartheta_1(\lambda)=\frac{1}{1+\tan^4(\frac{\theta}{2})\coth^2{(\lambda)}},~~\vartheta_2(\lambda)=\frac{1}{1+\tan^4(\frac{\theta}{2})\tanh^2{(\lambda)}}.
\label{thetai_XX}
\end{eqnarray}
Note that for the fully tilted state $\theta=\pi/2$,  these obey $\vartheta_2(\lambda)=1-\vartheta_1(\lambda)$, meaning they are interchanged under exchange of particles and  holes.  This relationship holds in general
\begin{eqnarray}
\vartheta_{p+1}(\lambda)=1-\vartheta_{p}(\lambda),~~\text{for}~\theta=\frac{\pi}{2}
\end{eqnarray}
 and reflects the spin flip symmetry of the fully tilted state.  As a consequence,  we find that for spin flip symmetric states,~\eqref{eq:qBT_1}-\eqref{eq:qBT_3} admit an exact solution
 \begin{eqnarray}
  q_{{\rm eff},j}&=&0,~ j<p\\
 q_{{\rm eff},p+1}&=& q_{{\rm eff},p}=\frac{p+1}{2}.
 \end{eqnarray}
From this one can appreciate the drastic effect the dressing operation on the magnetization of the strings. Its impact is most pronounced as we approach the isotropic point $p\gg 1$.  In that case,  the last two strings carry huge magnetization,  moreover, as the $p$ string is a bound state of $p$ spinons it extends over a large region and concomitantly becomes almost immobile.  As result of the $v_{p}=v_{p+1}$ this is also the case for the $p+1$ string and we have that the transport of magnetization from ballistic propagation of quasiparticles is suppressed. 
 
\textbf{Gapped regime:} In the gapped regime we instead parametrize the anisotropy by $\eta$ which is defined by $\Delta=\cosh(\eta)$ and unlike in the gapless regime we place no restrictions upon $\eta$.   The nature of the strings in this regime changes and as previewed above we have that $M=\infty$ and $\Lambda=\frac{\pi}{2}$.  The various distributions governing the stationary states of the model obey the integral equations 
\begin{eqnarray}\label{eq:BT_1_gapped}
\rho^{t}_1&=&s + s\star\left[(1-\vartheta_2)\rho^t_{2}+\delta_{p,2} \rho_{p+1}\right]\\\label{eq:BT_2_gapped}
    \rho^t_m &=&s\star\left[(1-\vartheta_{m-1})\rho^t_{m-1}+(1-\vartheta_{m+1})\rho^t_{m+1}\right]
\end{eqnarray} 
where now $s(\lambda)=\frac{1}{2\pi}\sum_{k\in\mathbb{Z}}e^{-2ik\lambda}\sech(k\eta)$.  Likewise, we have 
\begin{eqnarray}\label{eq:vBT_1_gapped}
v_1\rho^{t}_1&=&-\frac{\sinh\eta}{2} \partial_\lambda s
    + s\star\left[(1-\vartheta_2)v_2\rho^t_{2}\right]\\\label{eq:vBT_2_gapped}
    v_m \rho^t_m&=&s\star\left[(1-\vartheta_{m-1})v_{m-1}\rho^t_{m-1}+(1-\vartheta_{m+1})v_{m+1}\rho_{m+1}^t\right]
\end{eqnarray}
and 
\begin{eqnarray}\label{eq:qBT_1_gapped}
q_{{\rm eff},1}&=& s\star\left[(1-\vartheta_2)q_{{\rm eff},2}\right]\\\label{eq:qBT_1_gapped}
q_{{\rm eff},m}&=&s\star\left[(1-\vartheta_{m-1})q_{{\rm eff},m-1}+(1-\vartheta_{m+1})q_{{\rm eff},m+1}\right]\\\label{eq:qBT_3_gapped}
    \lim_{m\to\infty}q_{{\rm eff},m+1}&=&q_{{\rm eff},m}+1
\end{eqnarray}
As in the gapped case the occupation functions can be determined exactly using the integrability of the initial state ~\cite{piroli2016exact}.  However, it is sometimes more practical to instead obtain them through the TBA equations
\begin{eqnarray}
\log\left[\frac{1-\vartheta_1}{\vartheta_1}\right]&=&d(\lambda)-s\star\log[\vartheta_{2}]\\
\log\left[\frac{1-\vartheta_m}{\vartheta_m}\right]&=&d(\lambda)-s\star\log[\vartheta_{m-1}\vartheta_{m+1}]\\
\lim_{m\to\infty}\log\left[\frac{1-\vartheta_{m+1}}{\vartheta_{m+1}}\right]&=&\log\left[\frac{1-\vartheta_m}{\vartheta_m}\right]-\log\left[ \tan^4(\frac{\theta}{2})\right]
\end{eqnarray}
where the driving term is 
\begin{eqnarray}
d(\lambda)=\sum_{k\in \mathbb{Z}} e^{-2ik\lambda}\left[1-(-1)^k\right]\frac{\tanh(k \eta)}{k}. 
\end{eqnarray}
In practice we numerically evaluate the above integral equations by truncating at some large number of strings.  The accuracy is then checked against the analytic solution in the case of $\vartheta_m$ for small $m$ or known quantities such as the net magnetization.


\section{Entanglement asymmetry}
\label{sec:entangelementasymmetry}

To characterize the initial symmetry breaking in the system and its eventual dynamical restoration we employ the entanglement asymmetry.  This quantity measures the amount of symmetry breaking in a subsystem, $A$,  of a many-body quantum state by comparing the actual reduced density matrix of $A$ at a time $t$,  to a symmetrized version which averages the state over the action of the broken symmetry group, $G$.   While there are many metrics one could choose to make this comparison, the entanglement  asymmetry is defined as the quantum relative entropy between the aforementioned states~\cite{ares2023entanglement}
\begin{eqnarray}\label{eq:asymm_def}
\Delta S_A(t)=\tr[\rho_A(t)\{\log\rho_A(t)-\log \rho_{A,G}(t)\}]. 
\end{eqnarray}
Here $\rho_A(t)=\tr_{\bar A}[\ketbra{\Psi(t)}{\Psi(t)}]$ is the reduced density matrix of $A$ whose complement is $\bar A$ and,  $\rho_{A,G}(t)$ is its symmetrized version.  We denote by $\ell$, the size of the subsystem which we assume to be much smaller than the full system size, $\ell\ll L$.  We are interested in the behaviour in the thermodynamic and ballistic scaling limits and so take $L,\ell,t\to\infty$ with $\
\zeta=t/\ell$ held fixed. To define $\rho_{A,G}(t)$ we employ a more general form introduced in~\cite{klobas2024translation}.  For continuous groups it is given by
\begin{equation}\label{eq:symm_state}
    \rho_{A,G}(t)=\frac{1}{\text{Vol}(G)}\int_{ G} {\rm d} g\,\tr_{\bar A}\left[e^{-iH t }g\ketbra{\Psi(0)}{\Psi(0)}g^{-1}e^{i Ht}\right]
\end{equation}
where the integral is over representations of the group elements with $\mathrm{d}g$ being the Haar measure on $G$ and $\text{Vol}(G)$ is the volume of the group manifold.  This definition differs somewhat from the original one but offers several advantages.  The major difference is that the group elements act on the full system at $t=0$ in contrast to acting only on the subsystem at time $t$.  This modification allows one to consider the groups that do not act locally or which do not commute with the time evolution operator.  In the current formulation, the asymmetry quantifies how much the action of the group perturbs the unitary dynamics of the system. 
If the system is eventually expected to restore the broken symmetry at some late time then one can expect both $\rho_A(t)$ and $\rho_{A,G}(t)$ to approach the same late time value.  Hence, through the asymmetry we can understand how the symmetry breaking affects the eventual relaxation of the system.  Note that because of the properties of the relative entropy,  $ \Delta S_A(t)\geq 0$ with the lower bound being saturated for  $\rho_A(t)=\rho_{A,G}(t)$ signifying the restoration of symmetry.  Similar asymmetry measures have also been studied in quantum information settings~\cite{marvian2014extending}.

Specializing to the case at hand, we take $G=U(1)$ with groups element given by $g=e^{2i\alpha S^z},~\alpha\in(-\pi,\pi]$.  Since this commutes with the Hamiltonian and moreover can be written as a direct sum of operators acting on $A$ and $\bar A$, $ S^z=S^z_{A}\oplus S^{z}_{\bar A}$,  we can reduce~\eqref{eq:symm_state} to 
\begin{eqnarray}\label{eq:Symm_state_old}
 \rho_{A,S^z}(t)&=&\int_{-\pi}^\pi\frac{{\rm d}\beta}{2\pi} e^{2i\beta S^z_A}\rho_A(t)e^{-2i\beta S_A^z}\\\label{eq:symm_state_proj}
&=&\sum_{q=-\ell}^{\ell}\Pi_q \rho_A(t)\Pi_q.
\end{eqnarray}
This is the standard definition of the symmetrized  state.  In the second line we have introduced $\Pi_q$ which is the projector onto the eigenspace of $S^z_A$ with eigenvalue $q/2$ and in doing so used its Fourier representation
\begin{eqnarray}
\Pi_q=\int_{-\pi}^\pi\frac{{\rm d}\beta}{2\pi} e^{i(2S^z_A-q)\beta}. 
\end{eqnarray}
Note that when the generator does not have a decomposition into operators acting on $A$ and $\bar A$~\eqref{eq:symm_state} and~\eqref{eq:Symm_state_old} are not equivalent.  Depending on the symmetry generator however, such differences could become negligible in the thermodynamic limit~\cite{khor2023confinement}.  Using~\eqref{eq:symm_state_proj}, the asymmetry measures the difference between the reduced state and one which has been symmetrized at time $t$ rather than in the initial state.  The advantage of the latter is that it allows one to reduce \eqref{eq:asymm_def} to a convenient form using the properties of the projectors,
\begin{eqnarray}
\Delta S_A(t)=S[\rho_{A,S^z}(t)]-S[\rho_{A}(t)]~; ~~S[\rho]=-\tr[\rho\log\rho].
\end{eqnarray}
Thus $\Delta S_A(t)$ becomes a difference of von Neumann entanglement entropies calculated using the two states.  Naively, this places an upper bound on the asymmetry which scales as $\ell$, however $\rho_{A}(t)$ and $\rho_{A,S^z}(t)$ are not independent and in fact the asymmetry scales at most logarithmically, $\sim\log \ell$~\cite{capizzi2023universal,ares2023entanglement}. 

\textbf{Charged Moments:} In order to calculate $\Delta S_A(t)$ we use the replica trick. Introducing the R\'enyi entropy  $S^{(n)}[\rho]=\frac{1}{1-n}\log \tr[\rho^n]$ which reduces to $S[\rho]$ in the limit $n\to 1$ we have that~\cite{ares2023entanglement}
\begin{eqnarray}
\Delta S_A(t)&=&\lim_{n\to 1}\frac{1}{1-n}\log \frac{\tr [\rho^n_{A,S^z}(t) ]}{\tr[\rho^n_{A}(t)]}.
\end{eqnarray}
We then rewrite this expression in terms of objects known as the charged moments which are defined as 
\begin{eqnarray}
Z_n(\boldsymbol{\beta},t)=\tr[e^{2i\beta_1 S^z_A}\rho_A(t)e^{2i\beta_2 S^z_A}\rho_A(t)\dots e^{2i\beta_n S^z_A}\rho_A(t)].
\end{eqnarray}
where $\boldsymbol{\beta}=(\beta_1,\dots\beta_n)$.  Using which, we obtain
\begin{eqnarray}\label{eq:renyi_asymmetry}
\Delta S_A(t)&=&\lim_{n\to 1}\frac{1}{1-n}\log\sum_{q=-\ell}^\ell \prod_{j=1}^n\left(\int _{-\pi}^\pi\frac{{\rm d}{\beta_j}}{2\pi}e^{-i\beta_j q}\right)e^{-\ell \mathcal{F}_n(\boldsymbol{\beta},t)}.
\end{eqnarray}
Here we introduced the free energy density associated to the ratio of charge moments,  $\mathcal{F}_n(\boldsymbol{\beta},t)=\ell^{-1}\log\frac{Z_n(\boldsymbol{0},t)}{Z_n(\boldsymbol{\beta},t)}$.    

\textbf{Space-time duality:}
The reduction of the entanglement asymmetry to the charged moments is particularly advantageous as there exist a number of methods by which they can be calculated. The most widely applicable of these is the space-time duality approach to quantum dynamics. This method maps non-equilibrium quantities like $Z_n(\boldsymbol{\alpha},t)$ into a dual quantity which is instead calculated in an equilibrium system but where the roles of space and time have been swapped. The method can be applied quite generally~\cite{banuls2009matrix, bertini2018exact,bertini2019entanglement,ippoliti2021postselectionfree, lu2021spacetime, garratt2021manybody, lerose2021influence} but is particularly useful for integrable models were it provides analytic predictions for the charged moments~\cite{bertini2022growth,bertini2023nonequilibrium,bertini2024dynamics,rylands2024full,rylands2024microscopic}. In the latter case it builds upon earlier work on  the Loschmidt echo in integrable models~\cite{pozsgay2013dynamical,piroli2017from,piroli2018nonanalytic,rylands2019loschmidt}.

While space-time duality itself is quite intuitive the calculations to obtain predictions for charged moments can become somewhat involved, particularly in the case of a symmetry broken initial state.  The details are explained fully in~\cite{bertini2024dynamics,rylands2024microscopic} and we quote here only the general result, specializing to the gapped and gapless cases in subsequent sections.   Moreover, we only need to consider the region $n\approx 1$ since we are interested in taking the replica limit.  In that case,  the free energy density is approximately factorized amongst the replicas at all times (In the initial state the factorization is immediate since $\ket{\Psi(0)}$ is a product state),  
\begin{eqnarray}
\mathcal{F}_n(\boldsymbol{\beta},t)&=&\sum_{j=1}^n
 \mathcal{F}_1(\beta_j,t),
\end{eqnarray}
where the single replica quantities $ \mathcal{F}_1(\beta_j,t)$ depend on the occupation functions $\vartheta_m(\lambda)$.  It should be noted that although in the  interacting case the replica factorization is only approximate near $n\approx 1$,  it is exact for free models~\cite{ares2022entanglement}.  This factorization allows us to reduce the asymmetry to a more convenient form for analysis. Introducing the quantity,
\begin{eqnarray}
J_q(t)=\int _{-\pi}^\pi\frac{{\rm d}{\beta}}{2\pi}e^{-i\beta q-\ell \mathcal{F}_1(\beta,t)}
\end{eqnarray}
we can rewrite the asymmetry as~\cite{rylands2024microscopic},
\begin{eqnarray}\label{eq:J_asymmetry}
\Delta S_A(t)&=&\lim_{n\to 1}\frac{1}{1-n}\log\sum_{q\in\mathbb{Z}} [J_q(t)]^n\\
&=&-\sum_{q\in \mathbb{Z}}{\rm Re}[J_q(t)\log J_q(t).] 
\end{eqnarray}
where in the first line we have replaced $\sum_{q=-\ell}^\ell\to \sum_{q\in \mathbb{Z}}$.  Thus the behaviour of $\Delta S_A(t)$ can be understood through the simpler quantity $J_q(t)$ and in particular, its analysis can be broken down into two different cases.  In the first, provided $ \mathcal{F}_1(\beta,t)\gg 1/\ell$ we can evaluate the $\beta$ integral via a saddle point approximation 
\begin{eqnarray}\label{eq:J_short_time}
J_q(t)\approx\sum_{\beta^\star_q} \frac{1}{\sqrt{2\pi | \mathcal{F}''_1(\beta_q^\star,t)|}}e^{-iq\beta^\star_q-\ell  \mathcal{F}_1(\beta_q^\star,t)}
\end{eqnarray}
where $\beta^\star_q$ satisfy $ \mathcal{F}'_1(\beta_q^\star,t)=-iq/\ell$.  We will see below from the explicit forms of $ \mathcal{F}_1(\beta,t)$ that this is valid up to intermediate timescales.  Whereas in the second,  $ \mathcal{F}_1(\beta,t)\ll 1/\ell$ in which case
\begin{eqnarray}\label{eq:J_long_time}
J_{q}(t)\approx\delta_{q,0}-\ell \int _{-\pi}^\pi\frac{{\rm d}{\beta}}{2\pi} \mathcal{F}_1(\beta,t)e^{-i q\beta}
\end{eqnarray}
which will be valid at very large times.  The largest  contribution to~\eqref{eq:J_asymmetry} in this late time regime thus comes from $J_0(t)$ and to leading order one can express this as 
\begin{eqnarray}
\Delta S_A(t)\approx -{\rm Re}[(1-J_0(t))\log(1-J_0(t))],
\end{eqnarray}
and determine the late time behaviour purely from $J_0(t)$.
\section{Gapless Quench}
\label{sec:gapless}
Having laid the ground work we now come to the main part of this paper; the calculation of the entanglement asymmetry in the quenched XXZ model.  We split this into two sections, starting here with the gapless regime of the model.   For this we have that the single replica free energy density is given by~\cite{rylands2024microscopic}
\begin{equation}\label{eq:charge_moment_integrable_gapless}
 \mathcal{F}_1(\beta,t)=\frac{1}{2}\sum_{m=1}^{p+1}\nu_m\!\!\int_{-\infty}^\infty {\rm d}\lambda\, {\rm min}[2\zeta|v_m(\lambda)|-1,0]\rho^t_m(\lambda)\left(\mathcal{K}^{\beta}_{m}(\lambda)+\theta_m(\lambda)\log[x_m^{\beta}(\lambda)e^{2i\beta \nu_m q_m}]\right)
\end{equation}
where $\nu_{m<p+1}=1,~\nu_{p+1}=-1$, while 
\begin{eqnarray}
\mathcal{K}^{\beta}_{m}(\lambda)&=&\nu_m\log [1-\vartheta_m(\lambda)+\vartheta_m(\lambda)e^{-\log x_{m}^{\beta}(\lambda)}], 
\end{eqnarray} 
and $x^\beta_m(\lambda)$ obey a set of integral equations
\begin{eqnarray}\label{eq:xTBA_gapless_1}
\log x^\beta_m&=&s\star[\mathcal{K}^\beta_{m-1}+\mathcal{K}^\beta_{m+1}+\log [x^\beta_{m-1}x^\beta_{m+1}]+\delta_{m,p-1}\mathcal{K}^\beta_m],~~1\leq m< p\\\label{eq:xTBA_gapless_2}
\log x^\beta_p &=& \log x^\beta_{p+1}-2i\beta(p+1)=-i\beta (p+1)+s\star[\mathcal{K}^\beta_{p-1}+\log x^\beta_{p-1}]
\end{eqnarray}
and additionally are related to the dressed magnetization via
\begin{eqnarray}
\frac{{\rm d}\log[x^\beta_m(\lambda)]}{{\rm d}\beta} \Big |_{\beta=0}&=&-2i\nu_m\,q_{{\rm eff},m}.
\end{eqnarray}
From \eqref{eq:charge_moment_integrable_gapless}
 we can see that $\mathcal{F}_1(\beta,t)$ obeys a quasiparticle picture with the factor  min$[2\zeta|v_m(\lambda)|-1,0]$ counting the number of quasiparticle pairs which are fully contained within the subsystem~\cite{chalas2024multiple}.  In the $\zeta\to\infty$ limit,  all quasiparticle pairs must exit the subsystem meaning that $\lim_{\zeta\to\infty}\mathcal{F}_1(\beta,t)=0$ and consequently the asymmetry vanishes and the symmetry is restored. In the region of this long time limit we may use ~\eqref{eq:J_long_time}.  
 
In the opposite limit when a large enough number of quasiparticles remain in the subsystem we can avail of the saddle point approximation~\eqref{eq:J_short_time}. Combining that with the expressions given above we find that thr $J_q(t)$ are approximately normally distributed and the asymmetry simply becomes the Shannon entropy of this distribution ~\cite{bertini2024dynamics,rylands2024microscopic},
\begin{eqnarray}\label{eq:saddle_pt_asymm_1}
\Delta S_A(t)&=&\frac{1}{2}+\frac{1}{2}\log \ell \pi \chi(t),~~\\\label{eq:saddle_pt_asymm_2}
\chi(t)&=&\sum_{m=1}^{p+1}\int_{-\infty}^\infty {\rm d}\lambda\, {\rm max}[1-2\zeta|v_m(t)|,0][q_{{\rm eff},m}(\lambda)]^2\rho_m(\lambda)[1-\vartheta_m(\lambda)]. 
\end{eqnarray}
which can be used provided ${\rm max}[1-2\zeta|v_m(t)|,0]\gg 1/\ell$.  At $\zeta=0$ this expression should reproduce the initial state value which is, of course, independent of the model.  The decomposition into quasiparticle properties is highly nontrivial and the fact that it can reproduce the correct initial asymmetry is an important check on our results.  
\begin{figure}
\includegraphics[width=.45 \textwidth,trim=0 120 0 120 ,clip]{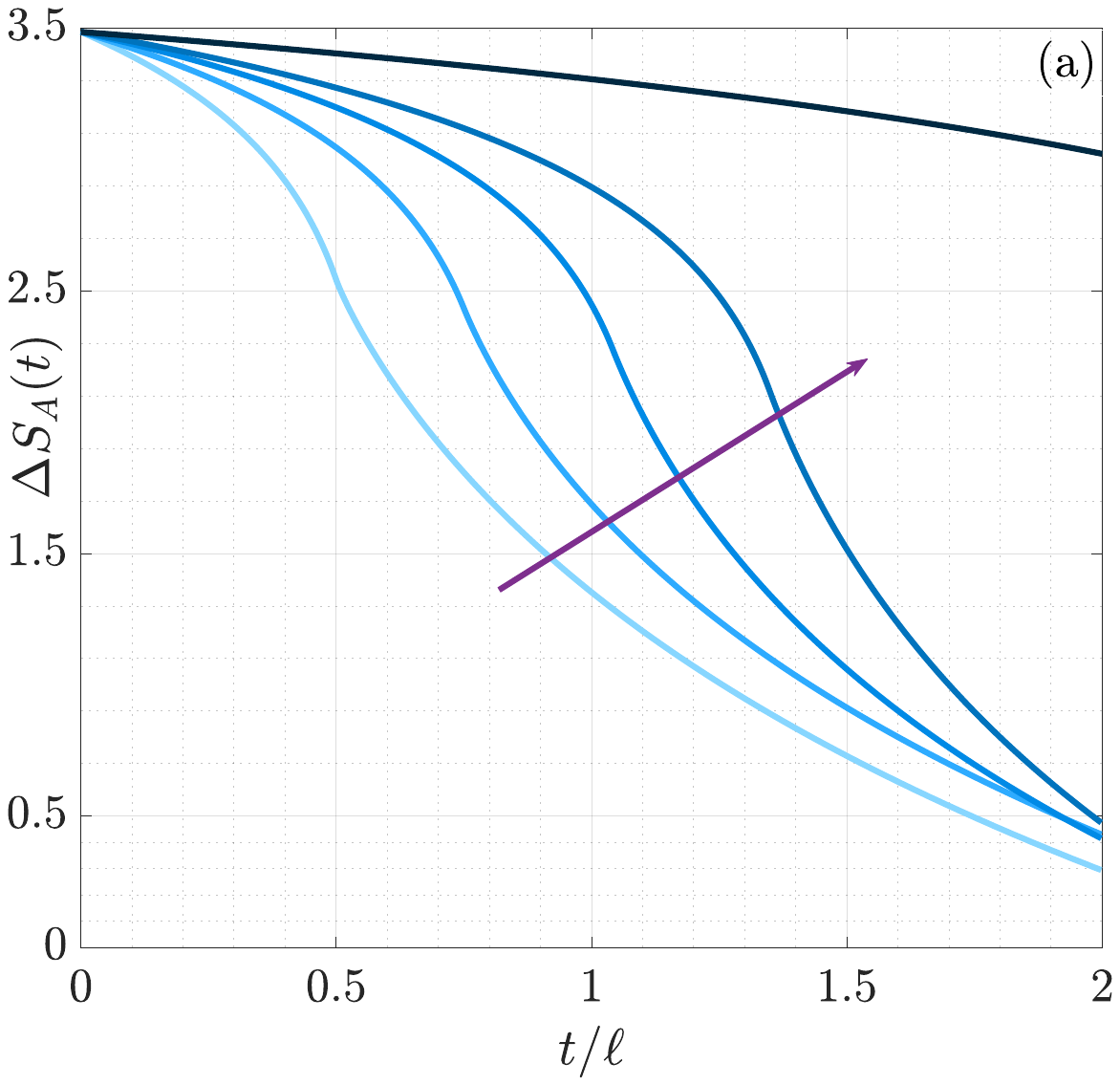}
\includegraphics[width=.45 \textwidth,trim=0 120 0 120 ,clip]{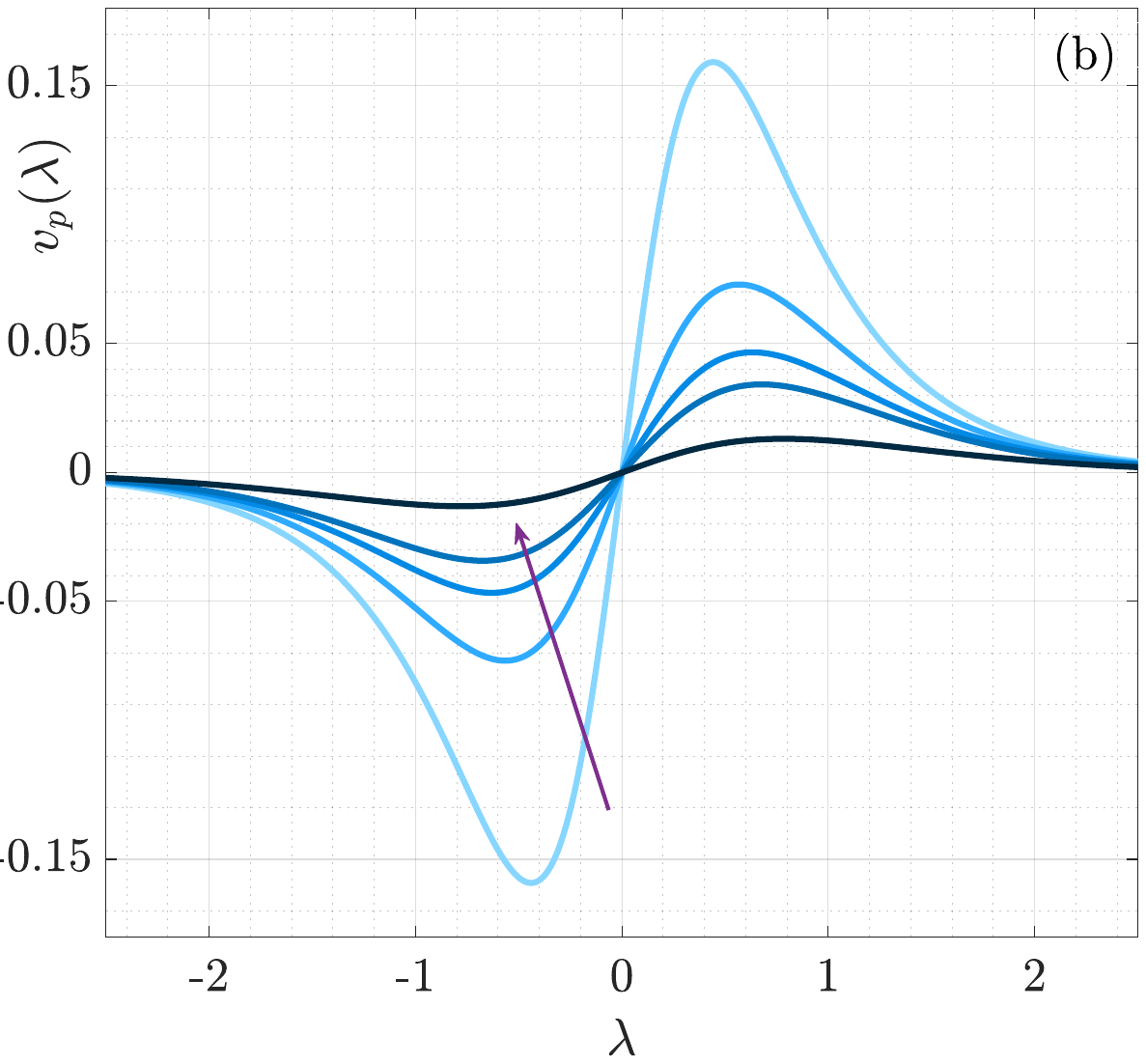}
\caption{(a) The entanglement asymmetry,  $\Delta S_A(t)$ as a function of $\zeta=t/\ell$ for subsystem size $\ell=1000$ quenched from the fully tilted state, $\theta=\frac{\pi}{2}$.    The different curves correspond to different values of $p=1,2,3,4,10$ going from light to darker, increasing along the arrow.  (b) The quasiparticle velocity $v_p(\lambda)$ for the fully tilted state for the same values of $p$, increasing along the arrow.     }\label{fig:fully_tilited_gapless}
\end{figure}

\textbf{Fully tilted state:}
A particularly instructive case study is given by the fully tilted state.  As mentioned above this state has a spin flip symmetry which is preserved by the dynamics.  Consequently,  only the last two strings are magnetized and moreover their contribution to the asymmetry is equal, 
\begin{eqnarray}
\chi(t)|_{\theta=\pi/2}=\frac{(p+1)^2}{2}\int_{-\infty}^\infty {\rm d}\lambda\,{\rm max}[1-2\zeta|v_p(\lambda)|,0] \rho_p(\lambda)[1-\vartheta_p(\lambda)].
\end{eqnarray}
In Fig.\ref{fig:fully_tilited_gapless}(a) we plot $\Delta S_A(t)$ using this expression for different values of anisotropy, $p=1,2,3,4,10$ as a function of $\zeta$, darker lines corresponding to larger $p$ and therefore increasing $\Delta$.  We see that as $p$ increases the dynamics slows down, the initial slope of the asymmetry becoming flatter and flatter until it is  almost constant near the isotropic point.  In (b) we show the velocities $v_p(\lambda)$ from which we see that they are suppressed for increasing $p$ (lighter to darker) explaining the slower symmetry restoration. 

We can also examine the asymmetry in the long time limit using~\eqref{eq:J_long_time}.  For this we can again use the extra spin flip symmetry to determine that
\begin{eqnarray}
\log[x^\beta_{p}(\lambda)]&=&- \log[x^\beta_{p+1}(\lambda)]=-i\beta(p+1)\\
\log[x^\beta_{m}(\lambda)]&=&0, ~~1\leq m <p
\end{eqnarray}
Using which we find that 
\begin{equation}\label{eq:J0_fully_tilted_gapless}
J_0(t)\approx\delta_{k,0}+\ell\int_{-\pi}^\pi\frac{{\rm d}\beta}{2\pi}\int_{-\infty}^\infty {\rm d}\lambda\, {\rm max}[1-2\zeta|v_p(\lambda)|,0]\rho^t_p(\lambda)\log[1-\vartheta_p(\lambda)+\vartheta_p(\lambda)e^{i\beta (p+1)}]e^{-iq\beta}. 
\end{equation}
To arrive at this result we have used the fact that $\mathcal{K}^\beta_{m<p}(\lambda)=0$ and that the contributions of the $p$ and $p+1$ strings are the same.  The form of~\eqref{eq:J0_fully_tilted_gapless} is the same as a free fermion model and as such its asymptotic behaviour can be studied following~\cite{rylands2024microscopic}.
We start by expanding the integrand as a power series in $e^{i\beta(p+1)}$, 
\begin{equation}
\log[1-\vartheta_p(\lambda)+\vartheta_p(\lambda)e^{-i\beta (p+1)}]={\rm max }[\log (\vartheta_p(\lambda)),\log (1-\vartheta_p(\lambda))]+\sum_{s=1}^\infty \frac{{\rm min}[\vartheta_p(\lambda),1-\vartheta_p(\lambda)]}{{\rm max}[\vartheta_p(\lambda),1-\vartheta_p(\lambda)]}e^{i\beta s(p+1)}. 
\end{equation}
Subbing this back into~\eqref{eq:J0_fully_tilted_gapless} we see immediately that at long time $J_{q}=0$ unless $q\geq 0$ and $q\equiv 0 \,{\rm mod}\,p+1$.  This contrasts with noninteracting case where only the odd $J_q$ vanish and is a result of the dressed value of the charge.  Retaining only the leading term using the fact that as $\zeta\to \infty$, ${\rm max}[1-2\zeta|v_p(\lambda)|,0]$ is only nonzero in a small window about $\lambda=0$ we find that the asymmetry vanishes at long time as 
\begin{eqnarray}
\Delta S_A(t)\approx c\ell \zeta^{-3}\log (\zeta). 
\end{eqnarray}
for some constant $c$.  

\begin{figure}
\includegraphics[width=.45\textwidth,trim=0 120 0 120 ,clip]{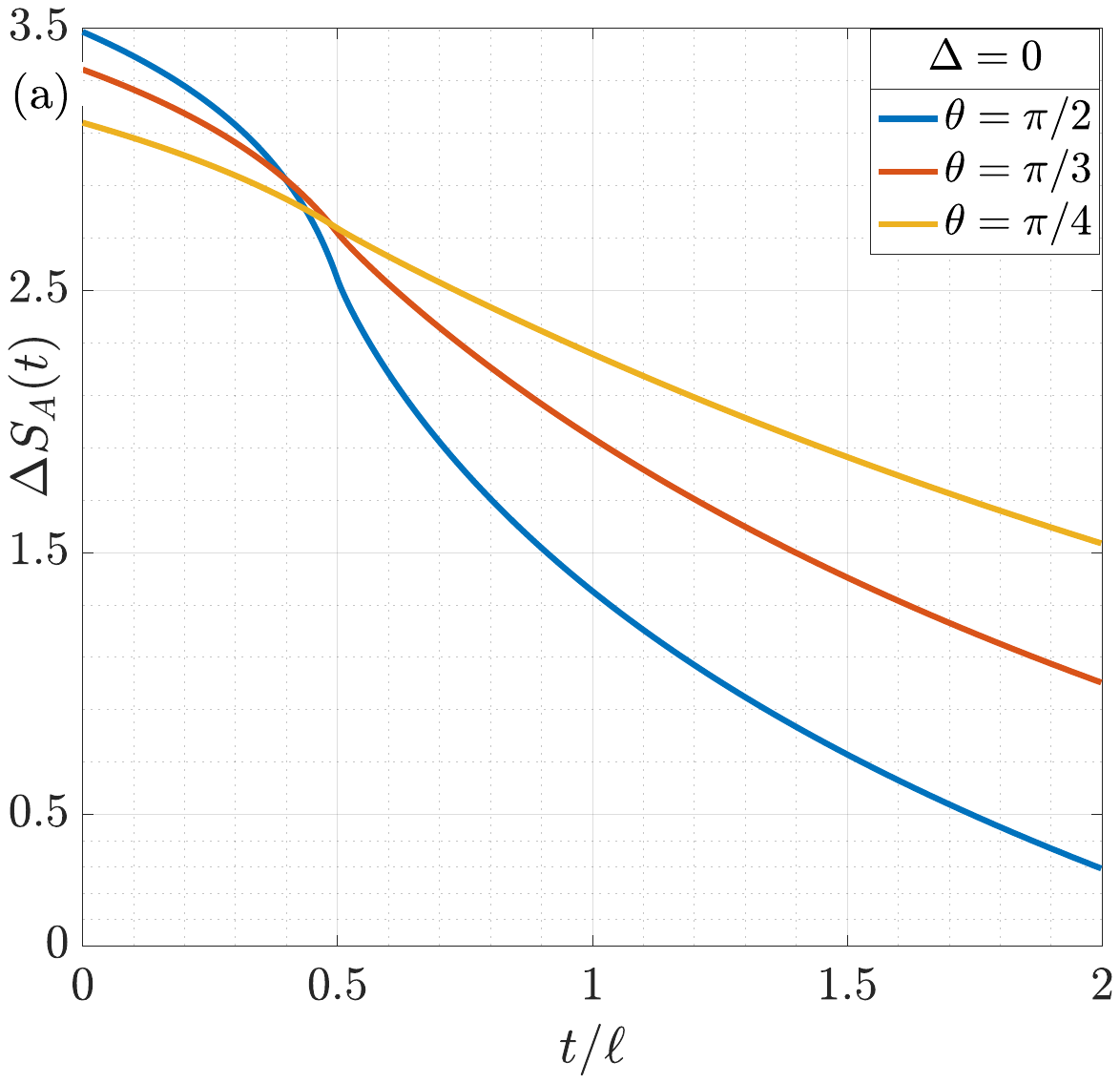}
\includegraphics[width=.45\textwidth,trim=0 120 0 120 ,clip]{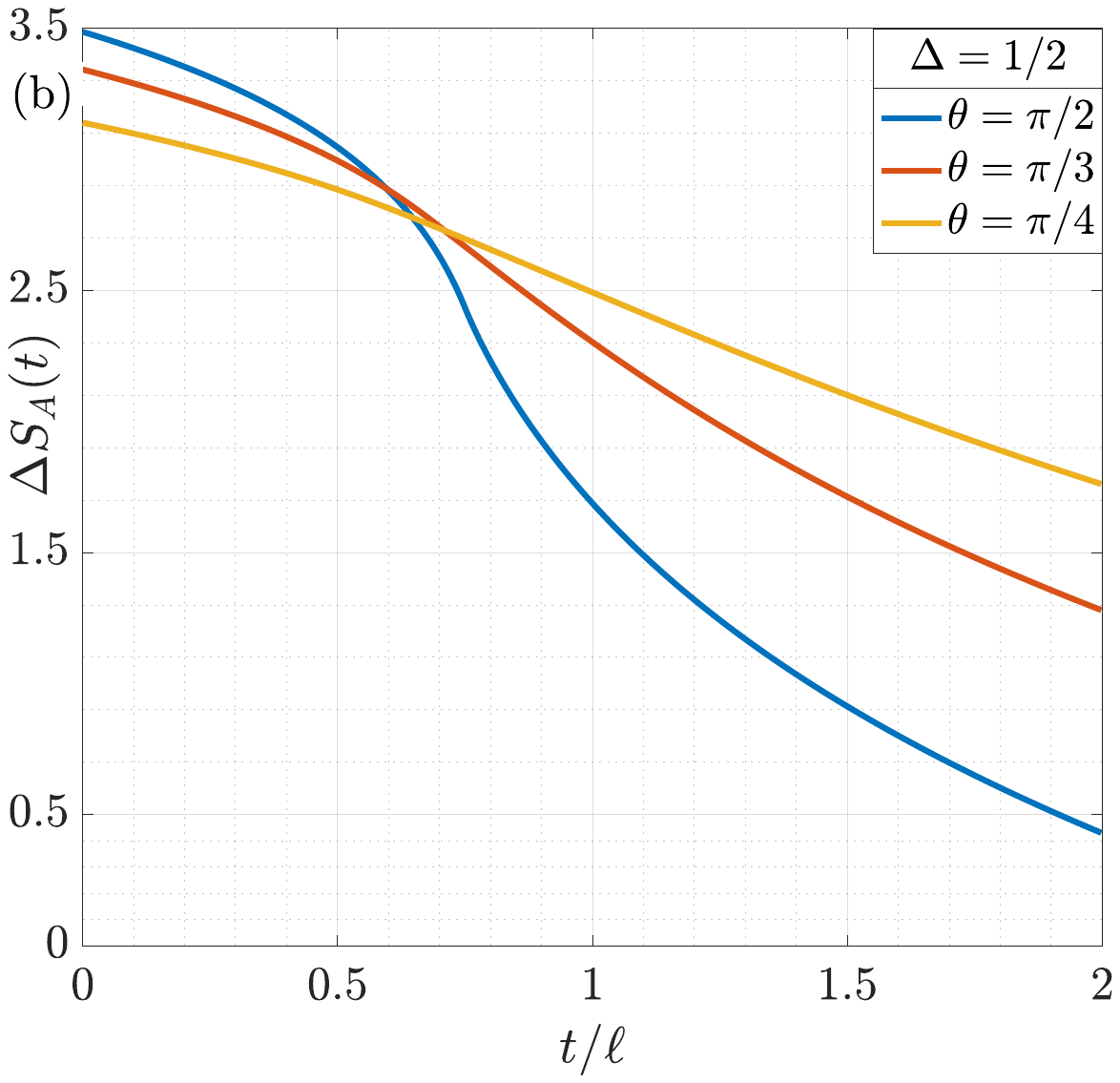}
\includegraphics[width=.45\textwidth,trim=0 120 0 120 ,clip]{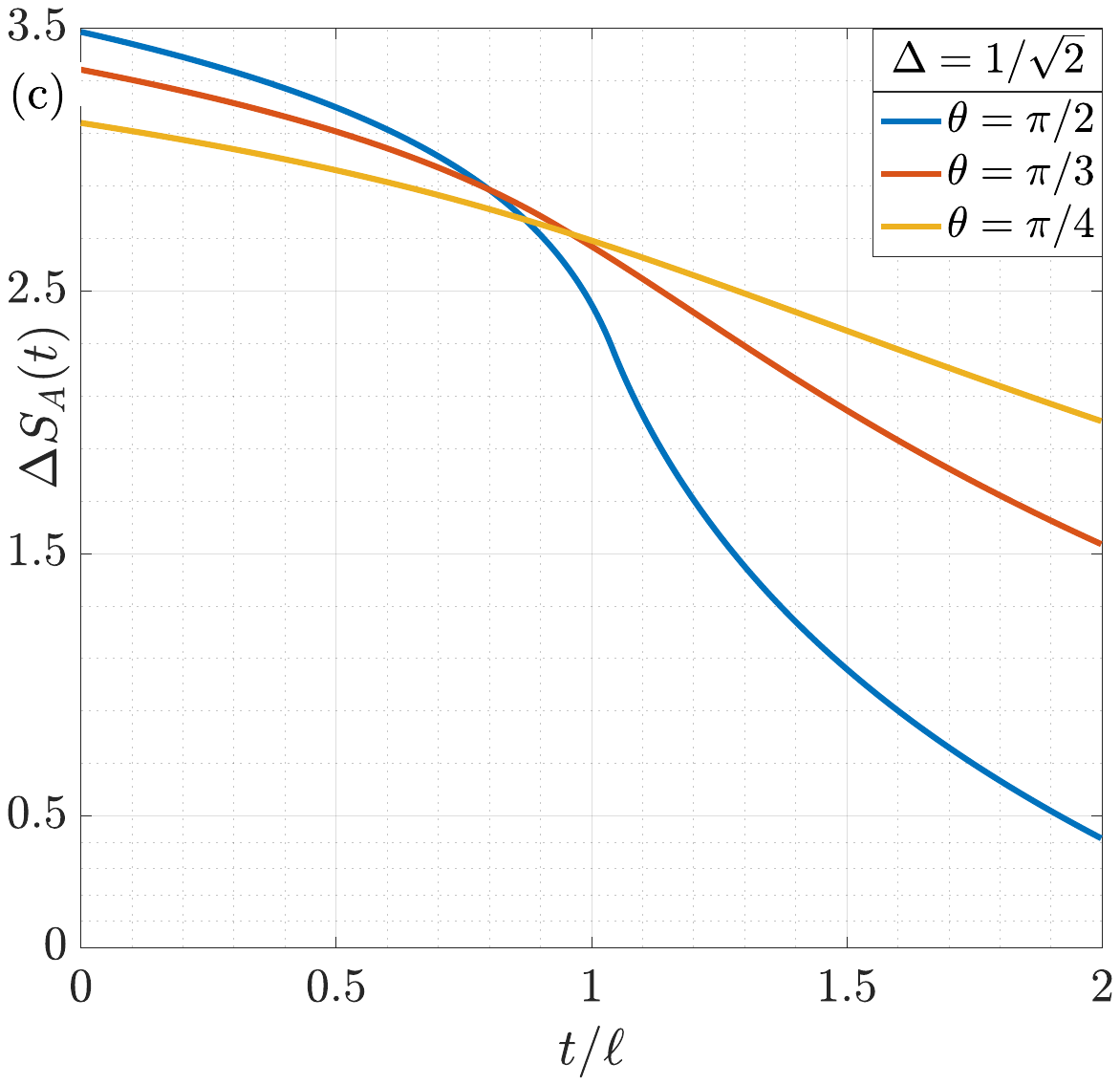}
\includegraphics[width=.45\textwidth,trim=0 120 0 120 ,clip]{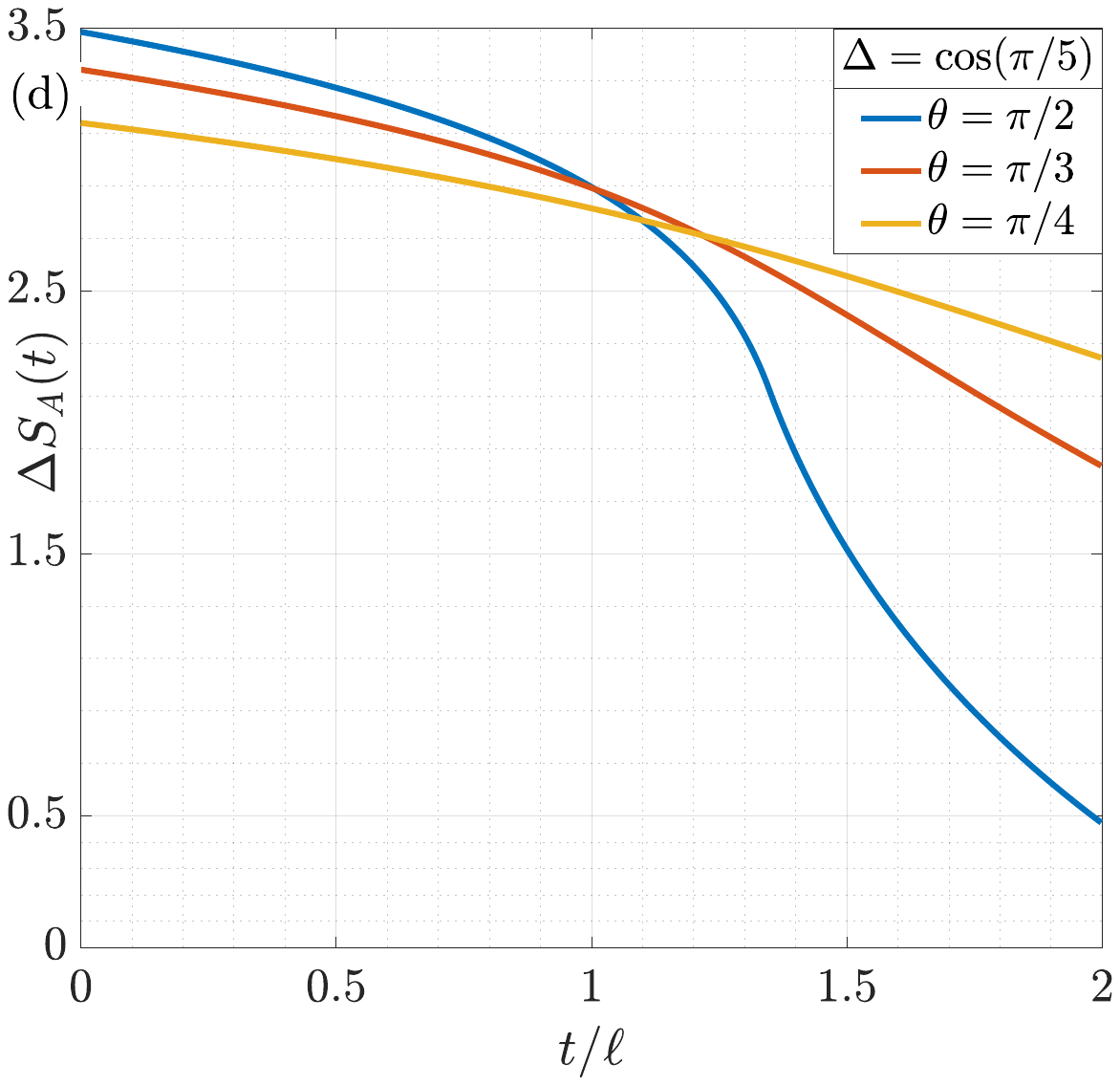}
\caption{The quantum Mpemba effect for different values of the anisotropy. We plot the entanglement asymmetry as function of $t/\ell$ for different initial tilt angles, $\theta=\pi/2,\pi/3,\pi/4$ and at different values of $p=1,\dots,4$ corresponding to (a)-(d) respectively.  We see the lines corresponding to different tilt angles cross in all cases indicating the quantum Mpemba effect. For increasing $\Delta$ or equivalently $p$ the time at which the crossing occurs increases.  }\label{fig:Mpemba_gapless}
\end{figure}

\textbf{Quantum Mpemba effect:}
Away from the fully tilted initial state, the system is less amenable to analytic treatment but can still be investigated by numerically integrating~\eqref{eq:saddle_pt_asymm_2}.   Upon doing so for different tilt angles  we can study the quantum Mpemba effect:  Given two states with tilt angles $\theta_1$ and $\theta_2$ and entanglement asymmetries $\Delta S_{A,1}(t)$ and  $\Delta S_{A,2}(t)$ respectively the quantum Mpemba effect occurs if 
\begin{eqnarray}
\Delta S_{A,1}(0)&>&\Delta S_{A,2}(0)\\
\Delta S_{A,1}(\tau)&<&\Delta S_{A,2}(\tau),~~\forall \tau >t_\text{M}
\end{eqnarray}
where $t_{\rm M}$ is some finite time,  called the Mpemba time.  The meaning is straightforward,  we have two states, one of which (state $1$),  is initially more asymmetric than the other.  They are allowed evolve unitarily under the same Hamiltonian and after a certain time $t_{\rm M}$ the other state (state 2) is more asymmetric and stays that way.  Thus the state which was initially more asymmetric restores the symmetry faster. If we use symmetry restoration as a coarse grained proxy for relaxation to the stationary state then we can infer that state 1 will relax faster than state 2.   In Fig.~\ref{fig:Mpemba_gapless} we plot~\eqref{eq:saddle_pt_asymm_1} for different values of the tilt angle, at $p=1,\dots,4$, (a)-(d) respectively.  We see that the quantum Mpemba effect occurs and is manifested by a crossing of the lines at finite time.  As $p$ increases the time at which this occurs also increases until the isotropic point where the lines remain flat as $\ket{\Psi}$ is an eigenstate of $H$ at $\Delta=1$.  

\begin{figure}
\includegraphics[width=.45 \textwidth,trim=0 120 0 120 ,clip]{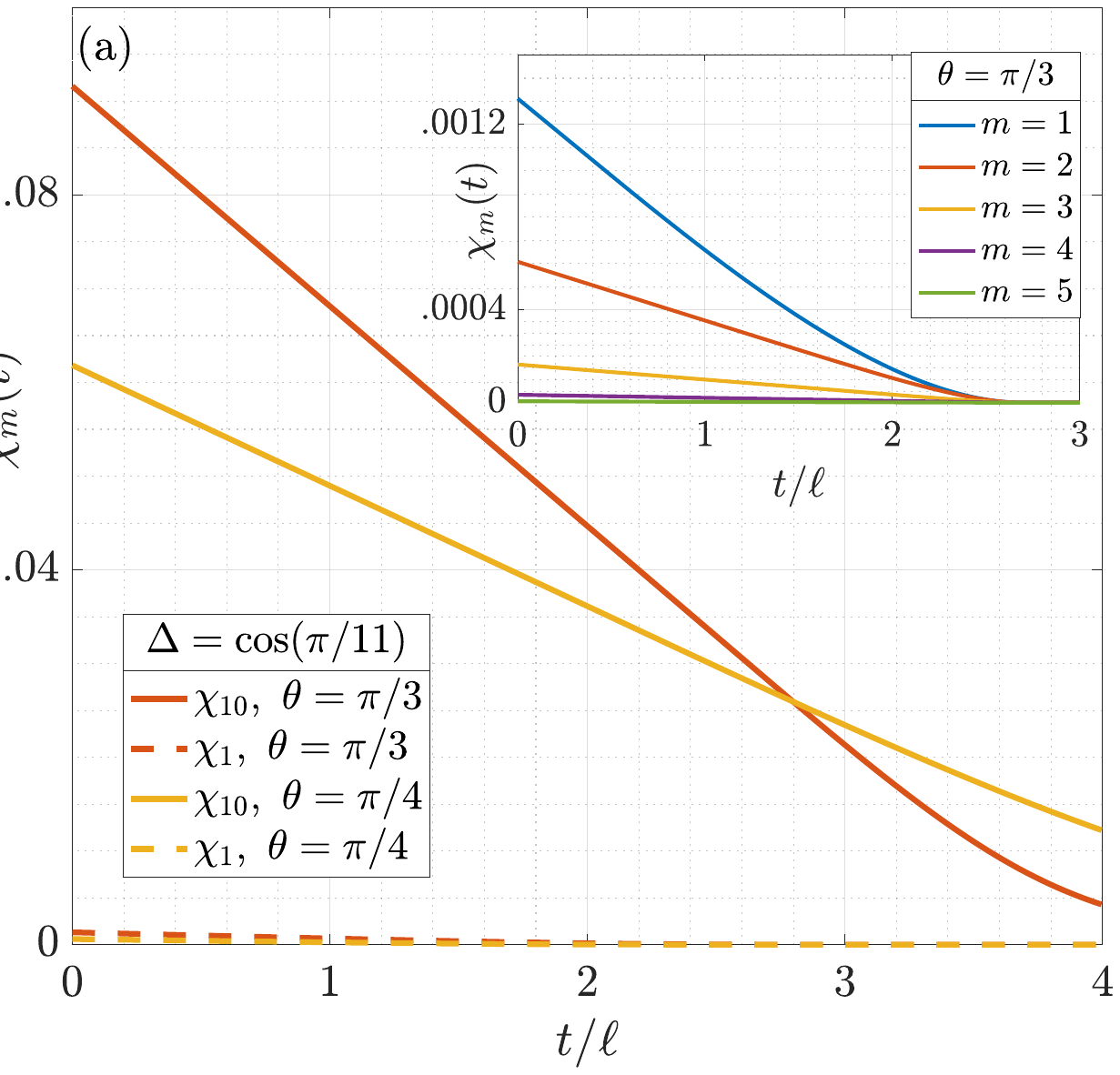}
\includegraphics[width=.45 \textwidth,trim=0 120 0 120 ,clip]{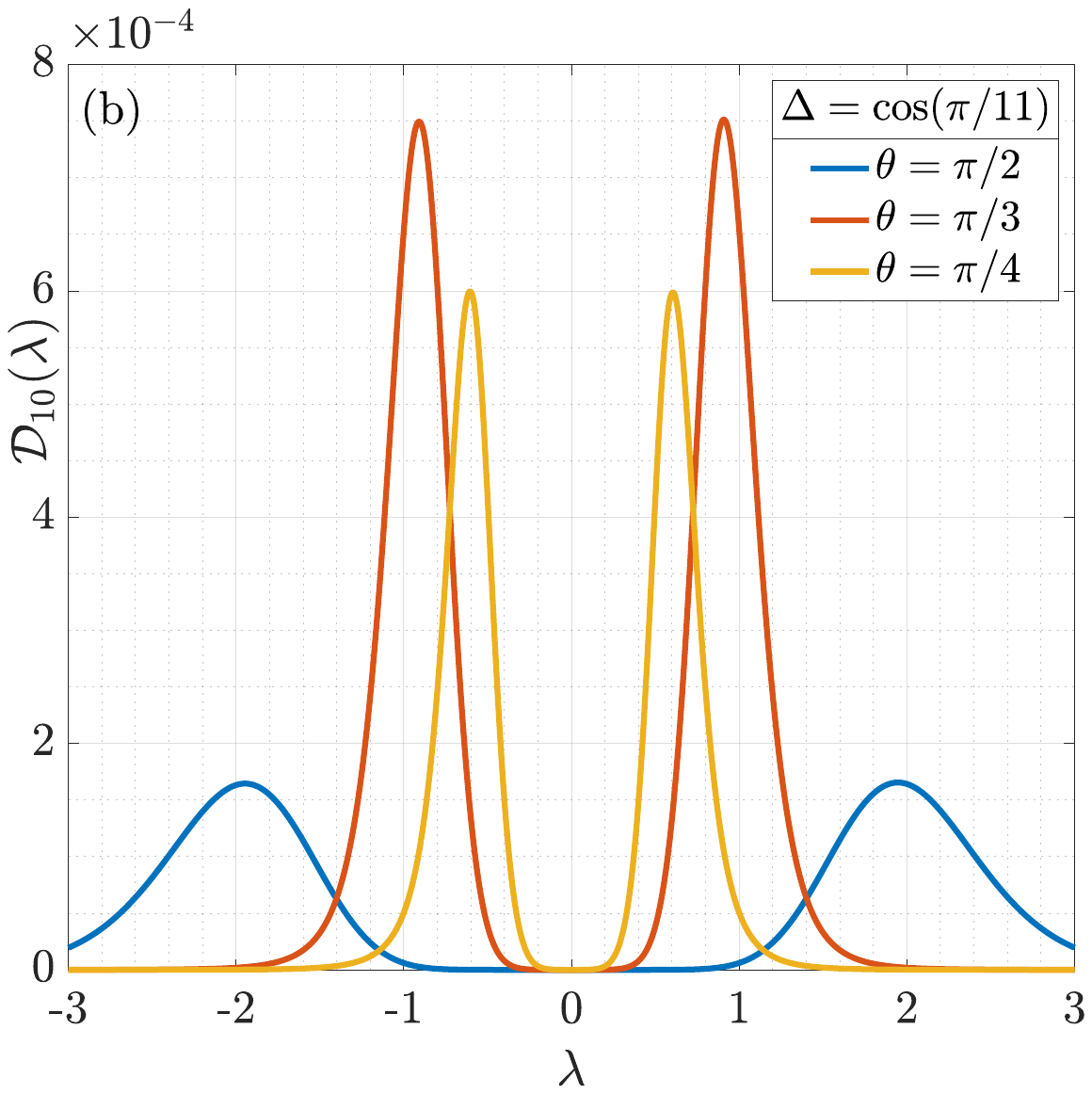}
\caption{(a) The quasiparticle contributions to $\chi(t)=\sum_{m=1}^{p+1}\chi_{m}(t)$ at $p=10$ for $m=1$ (dashed lines) and $m=10$ (solid). We plot the cases for $\theta=\pi/3,\pi/4$. In the inset we plot $\chi_m$ for $m=1,\dots,5$ for $\theta=\pi/3$. From this we see that the contributions of the second last string dominate all others.  (b) The quasiparticle Drude self weight for the $m=10$ string, $\mathcal{D}_{10}(\lambda)$, also for $p=10$ but with $\theta=\pi/2,\pi/3,\pi/4$.  Recall that in the fully tilted case $\mathcal{D}_{11}=\mathcal{D}_{10}$  while all other strings do not contribute.  For $\theta\neq\pi/2$ smaller strings have a non zero but negligible value of $\mathcal{D}_m$. }\label{fig:Drude}
\end{figure}

To investigate this behaviour we separate~\eqref{eq:saddle_pt_asymm_2}  into its contributions from different quasiparticle species, i.e. $\chi(t)=\sum_{m=1}^{p+1}\chi_m(t)$ and study which species contribute most.  In the fully tilted case we have seen that only $\chi_{m}(t)$ with $m=p,p+1$ are non zero and have the same contribution.  Away from this value, the longest string has the dominant contribution, $\chi_p(t)\gg\chi_m(t),~m\neq p$. We see this is the case in Fig.~\ref{fig:Drude} (a) where we plot the values of $\chi_m(t)$ for $\theta=\pi/3,\pi/4$ with $m=1$ (dashed lines) and $m=10$ (solid lines).  In the inset we see that all other species have smaller contributions than these two.  Thus we can understand the behaviour of $\Delta S_A(t)$ through the dynamics of the longest strings.  To this end, we examine the quasiparticle Drude self weight 
\begin{eqnarray}
\mathcal{D}_m(\lambda)=|v_m(\lambda)|[q_{{\rm eff},m}(\lambda)]^2\rho_m(\lambda)[1-\vartheta_m(\lambda)]
\end{eqnarray}
which governs the contribution of the quasiparticle $m$ at rapidity $\lambda$ to the slope of $\Delta S_A(t)$ given in~\eqref{eq:saddle_pt_asymm_1}.  This is also related to the connected two point function of the time integrated spin current through a point, $\mathcal{J}(t)$, via $\sum_{m=1}^M\int _{-\Lambda}^\Lambda\mathcal{D}_m(\lambda)=\left<\right.\![\mathcal{J}(t)]^2\!\left.\right>^{\rm c}$ and thus characterizes the transport of magnetization~\cite{doyon2017drude,myers2020transport}.  We plot this for the longest strings in Fig.~\ref{fig:Drude} (b) for different values of the tilt angle, $\theta=\pi/2,\pi/3,\pi/4$ at $p=10$, similar behaviour occurs at other values of $p$ also.  We see here that $\mathcal{D}_p(\lambda)$ has peaks which are symmetric about the origin and moreover these peaks shift inward as the tilt angle decreases. Thus as the tilt angle is decreased the spin Drude self weight is dominated by slower quasiparticles.  Accordingly, the transport of magnetization is also slower, resulting in the later restoration of the symmetry of smaller tilt angles witnessed in Fig. ~\ref{fig:Mpemba_gapless}. 

\section{Gapped Quench}
\label{sec:gapped}

We now turn our attention to the dynamics of the entanglement asymmetry in the gapped regime, $\Delta>1$.  Here the single replica free energy density is given by
\begin{figure}
\includegraphics[width=.45 \textwidth,trim=0 120 0 120 ,clip]{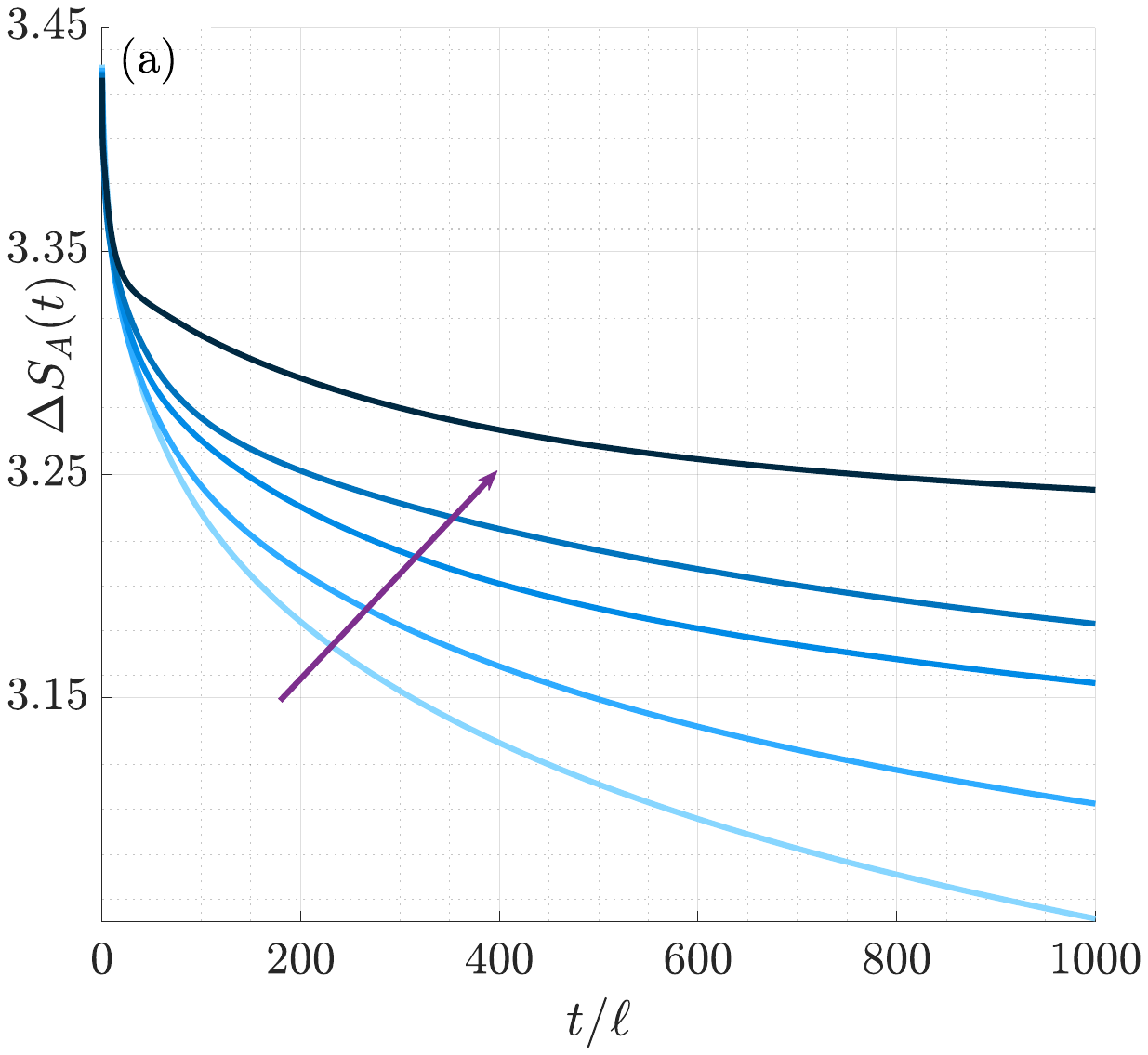}
\includegraphics[width=.45 \textwidth,trim=0 120 0 120 ,clip]{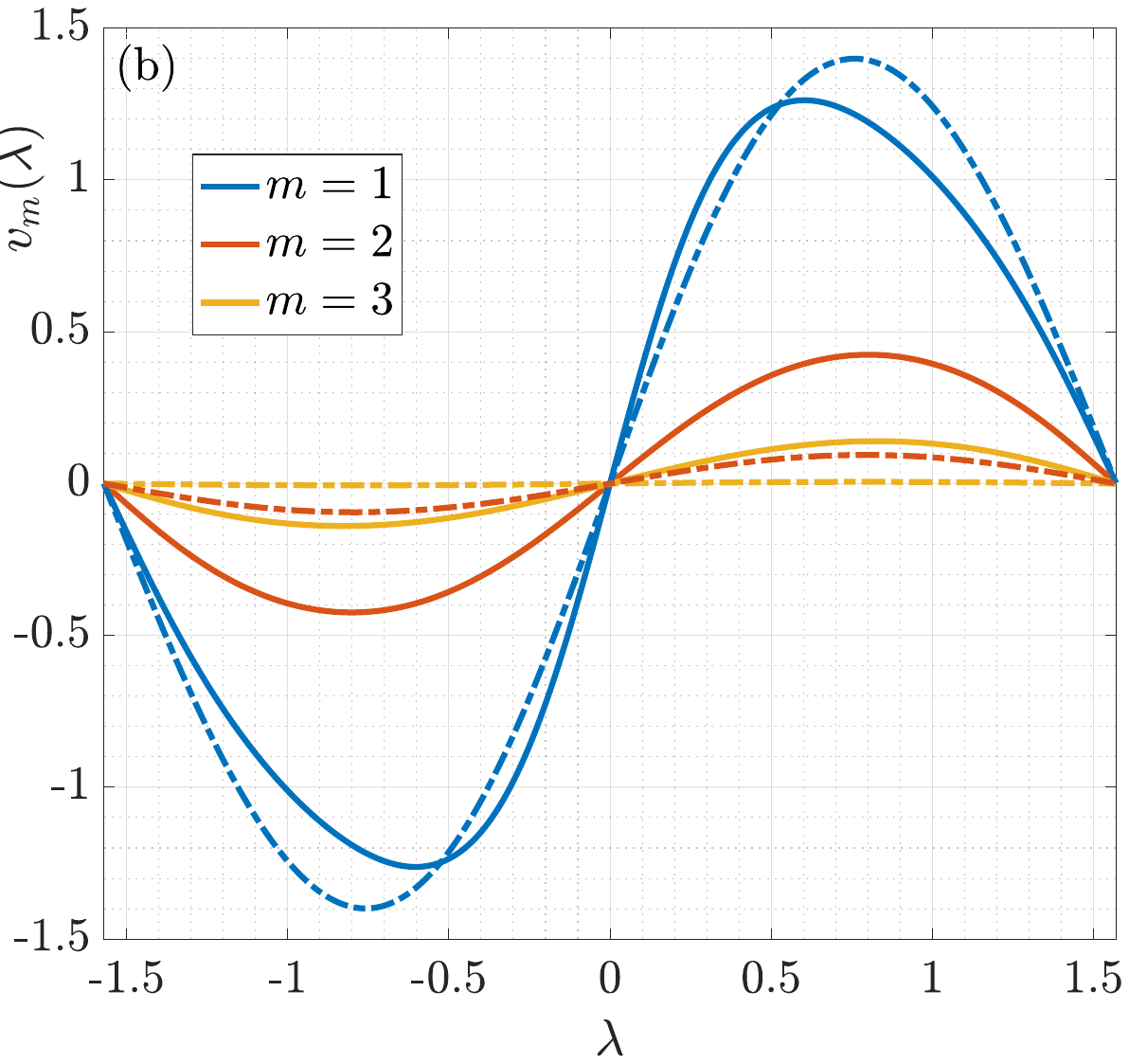}
\caption{(a)$\Delta S_A(t)$ for $\ell=1000$ at $\theta=2\pi/5$ with $\Delta=1.5,2,3,4,10$ (increasing lighter to darker along the arrow).  (b) The quasiparticle velocity $v_m(\lambda)$ for different $m=1,2,3$ and at $\Delta=2$ (solid lines) and $\Delta=10$ (dash-dotted lines).  
Both figures are obtained by truncating to a maximal string length of $30$. }\label{fig:gapped_5pi_2}
\end{figure}
\begin{eqnarray}\label{eq:charge_moment_integrable_pped}
 \mathcal{F}_1(\beta_j,t)&=&\frac{1}{2}\sum_{m=1}^\infty\int_{-\pi/2}^{\pi/2} {\rm d}\lambda\, {\rm min}[2\zeta|v_m(\lambda)|-1,0]\rho^t_m(\lambda)\left(\mathcal{K}^{\beta_j}_{m}(\lambda)+\theta_m(\lambda)\log[x_m^{\beta}(\lambda)e^{2i\beta_jq_m}]\right),\\
\mathcal{K}^{\beta}_{m}(\lambda)&=&\log [1-\vartheta_m(\lambda)+\vartheta_m(\lambda)e^{-\log x_{m}^{\beta}(\lambda)}]
\end{eqnarray}
with $x^\beta_m(\lambda)$ this time obeying,
\begin{eqnarray}
\log x^\beta_m&=&s\star[\mathcal{K}^\beta_{m-1}+\mathcal{K}^\beta_{m+1}+\log [x^\beta_{m-1}x^\beta_{m+1}]]\\
\lim_{m\to\infty} \log x^\beta_{m+1}&=&\log x^\beta_m-2i\beta.
\end{eqnarray}
As before, these can  be related to the dressed magnetization via
\begin{eqnarray}
\frac{{\rm d}\log[x^\beta_m(\lambda)]}{{\rm d}\beta} \Big |_{\beta=0}=-2iq_{{\rm eff},m}.
\end{eqnarray}
Unlike the gapless case, there are no special values of the initial tilt angle which allow for easier analysis. Indeed,  at $\theta=\pi/2$ our methods breakdown as they cannot detect sub-ballistic dynamics which is expected to occur there~\cite{znidaric2011spin,karrasch2017real,illievski2018superdiffusion,gopalakrishnan2019kinetic,bertini2021finite}. We therefore proceed directly to evaluating $\Delta S_A(t)$ in the saddle point regime,~\eqref{eq:saddle_pt_asymm_1} using 
\begin{eqnarray}
\chi(t)&=&\sum_{m=1}^{\infty}\int_{-\pi/2}^{\pi/2} {\rm d}\lambda\, {\rm max}[1-2\zeta|v_m(t)|,0][q_{{\rm eff},m}(\lambda)]^2\rho_m(\lambda)[1-\vartheta_m(\lambda)]. 
\end{eqnarray}
In Fig.~\ref{fig:gapped_5pi_2} (a) we plot $\Delta S_A(t)$ for $\theta=2\pi/5$ for different values of $\Delta=1.5-10$ increasing lighter to darker along the arrow.  Here we see a notable difference in the behaviour of the asymmetry, namely the dynamics is much slower, showing only a small decrease over 3 decades.  The relaxation becomes slower as $\Delta$ increases.   To understand this slower relaxation, in  Fig.~\ref{fig:gapped_5pi_2} (b) we plot the quasiparticle velocity $v_m(\lambda)$ for $m=1,2,3$ at $\Delta=2$ (solid lines) and $\Delta=10$ (dashed lines). From this, we can see that the velocity of a string is suppressed as its length increases, a feature which is accentuated as $\Delta$ increases.  Unlike in the gapless case however there is no drastic dressing of the magnetization each string carries. Therefore long strings still carry large magnetization in the gapless regime but their velocity is extremely suppressed. This leads to very slow spin dynamics and decay of the asymmetry.  The initial drop in the asymmetry, seen in Fig.~\ref{fig:gapped_5pi_2} (a) can therefore be attributed to the shortest strings, with the slowly decaying tail coming from the much longer strings.  

\begin{figure}
\includegraphics[width=.45 \textwidth,trim=0 120 0 120 ,clip]{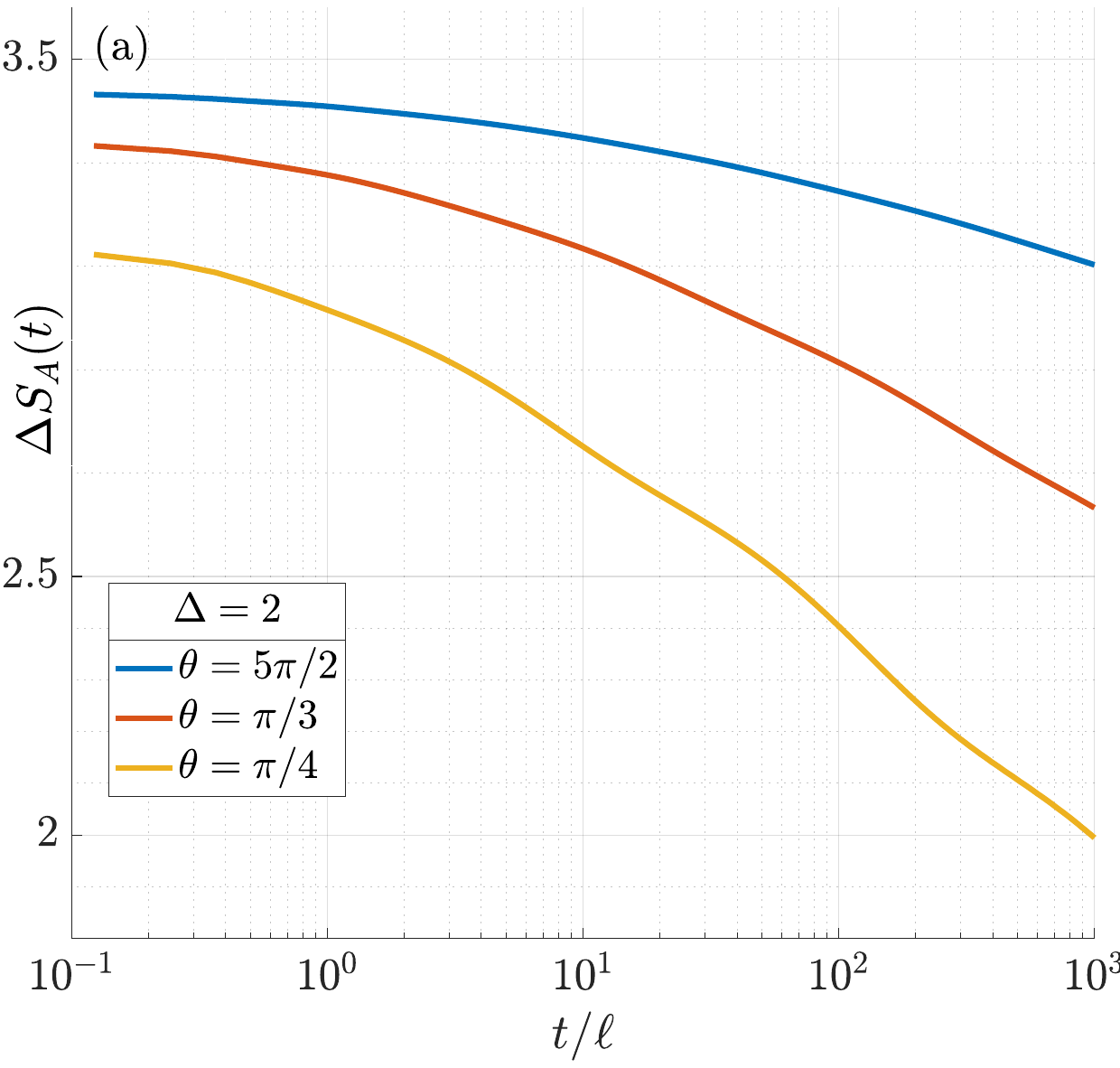}
\includegraphics[width=.45 \textwidth,trim=0 120 0 120 ,clip]{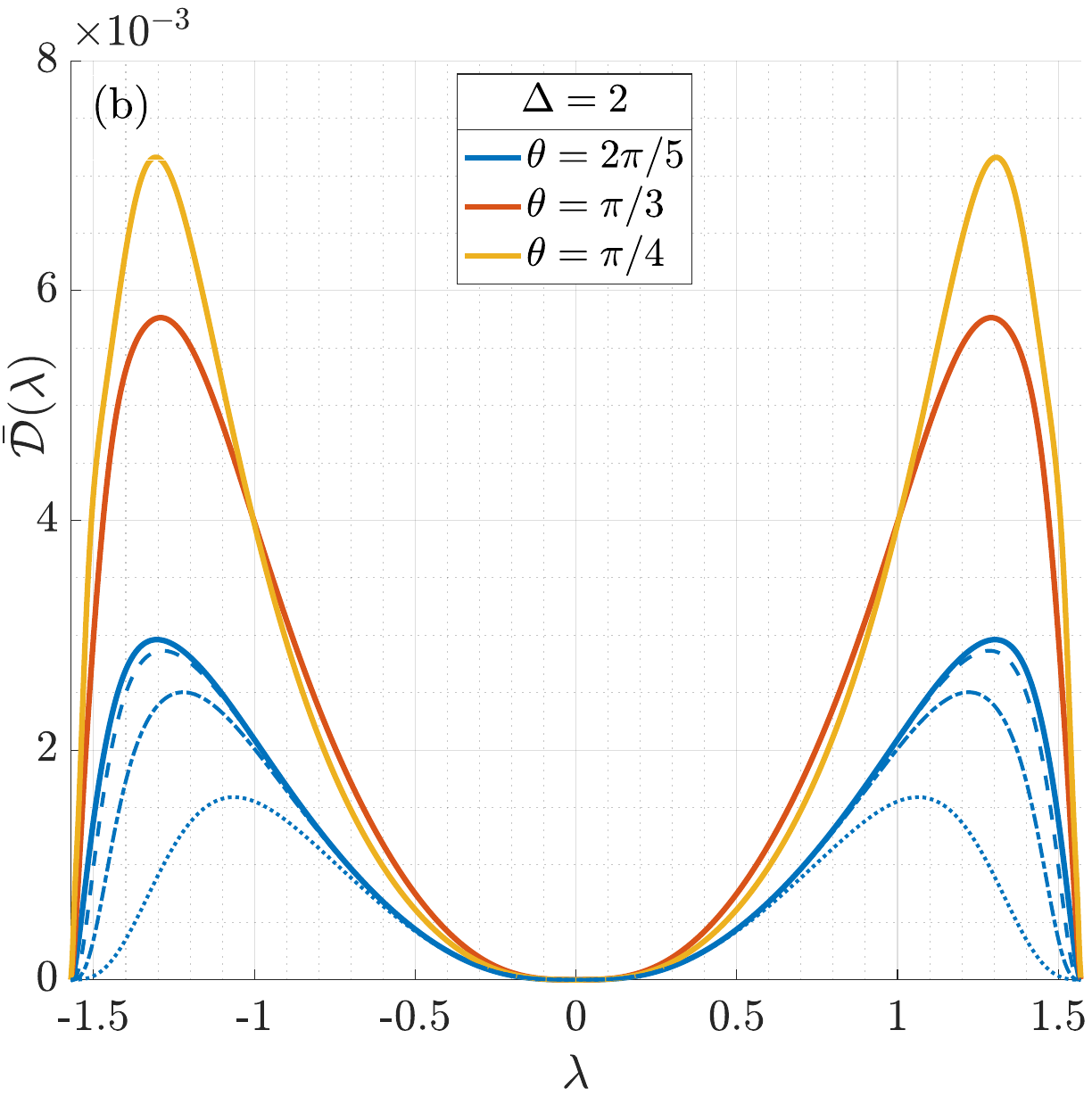}
\caption{(a) Linear-log plot of $\Delta S_A(t)$ for $\ell=1000$ at $\Delta=2$ and $\theta=2\pi/5,\theta/3,\theta/4$ as a function of $t/\ell$.  (b) The total Drude weight self weight $\bar{\mathcal{D}}(\lambda)=\sum_{m}\mathcal{D}_m(\lambda)$ as a function of $\lambda$ for same tilt angles (solid lines). We also plot $\mathcal{D}_1$ (dotted line) $\mathcal{D}_1+\mathcal{D}_2$ (dash-dotted) and $\mathcal{D}_1+\mathcal{D}_2+\mathcal{D}_3$ (dashed) for $\theta=2\pi/5$. From this we see that the smallest strings contribute the most to the spin transport. Both figures are obtained by truncating to a maximal string length of $30$. }\label{fig:asymm_Delta_2}
\end{figure}

Comparing $\Delta S_A(t)$ for different values of tilt angle we again find a distinct departure form the gapless case.  In Fig.~\ref{fig:asymm_Delta_2}(a) we plot $\Delta S_A(t)$ for $\Delta=2$ at $\theta=2\pi/5,\pi/3,\pi/4$ as a function of $t/\ell$ and see that no crossing between the different curves occurs,  a feature which persists for other values of $\theta$ and $\Delta$. This  means that the quantum Mpemba effect is absent.  To investigate this further we take a closer look at the quasiparticle Drude self weight per rapidity $\bar{\mathcal{D}}(\lambda)=\sum_m\mathcal{D}_m(\lambda)$ which governs the slope of $\chi(t)$ as well as  the spin transport in the system.   In Fig.~\ref{fig:asymm_Delta_2}(b) we plot this as a function of $\lambda$ for different tilt angles (solid lines).  As with the gapless regime this quantity exhibits a two peak structure,  in contrast however the peaks do not shift as the tilt angle is changed, thereby explaining the absence of Mpemba effect. Moreover, as shown by the dotted and dashed lines, the Drude self weight is actually dominated by the smallest strings despite the fact that they carry the least magnetization. This is a consequence of the diminished velocity of the longer strings which is not compensated for by their increased magnetization.  

\section{Conclusions}
\label{sec:conslusions}
In this paper we have investigated the dynamical restoration of $U(1)$ symmetry, corresponding to magnetization along the $z$ direction,  in the XXZ spin chain quenched from a tilted ferromagnetic state.  To do this we have calculated the dynamics of the entanglement asymmetry which measures how close the reduced density matrix of the system is to being symmetric.  Using this we find two distinct patterns of symmetry restoration depending on which regime the the model is in.  

In the gapless regime,  at roots of unity, the magnetization is predominantly transported by the longest possible strings, which are restricted to be of finite length.  This is captured by the Drude self weight, which governs spin transport in the model.  It is dominated by the contribution of the longest string with shorter strings having negligible contribution.  As the anisotropy is increased away from the free point towards the isotropic point the velocity of these maximal length strings decreases, leading to a suppression of the Drude self weight and slower restoration of the symmetry.  This behaviour is exemplified for an initially fully tilted state, wherein the magnetization of the first $p-1$ strings is zero due to spin-flip symmetry and the asymmetry can be written in terms of the longest string.  Comparing different tilt angles one can witness the quantum Mpemba effect-- the larger the initial tilt angle, the faster the symmetry is restored.  This is reflected in the Drude self weight,  which is peaked about slower and slower modes as the tilt angle decreases.  

In the gapped case the situation changes.  Here, the time required to restore the symmetry is several orders of magnitude longer than in the gapped case.  We can understand this from the nature of the quasiparticles in the gapped regime.  Therein there is no restriction on the length of the string and also no drastic dressing of their magnetization, meaning that the magentization is carried by exceedingly large bound states of spinons.   However, the velocity of these strings is significantly diminished with increasing string length.  This is reflected in the Drude self weight which, in contrast to the gapless case,  is dominated by the shortest strings.  The asymmetry therefore experiences an initial decrease due to the short strings, followed by a long slowly decaying tail coming from the longer strings.   If we now compare different  initial tilt angles the quantum Mpemba effect no longer occurs.  Once again, the Drude self weight provides some insight.  As in the gapless case, this exhibits a two peak structure however, unlike the gapless case, the position of these peaks does not shift as the tilt angle is changed.  Accordingly, the restoration of the symmetry is not significantly affected by altering the tilt angle.  In the limit of a fully tilted initial state, the system has diffusive spin transport and therefore at ballistic scales the asymmetry remains constant.  For any smaller tilt, the transport is instead ballistic, although slow, and so it is not possible for the maximally tilted state to restore the symmetry first. 

 This paper presents the first systematic calculation of the entanglement asymmetry across the full range of a parameter in an interacting integrable model. Remarkably, the occurrence of the Mpemba effect closely mirrors the zero-temperature phase diagram of the model, despite the fact that the physics is governed by high-energy eigenstates. It would be intriguing to explore whether a similar phenomenon arises in more complex spin chains, such as those with a nested Bethe ansatz structure, whose out-of-equilibrium dynamics have been solved in Refs. \cite{Mestyan_2017,Piroli_2019,Piroli_2019b}.

\begin{acknowledgements}
    CR and PC wish to thank F. Ares, B. Bertini, K. Klobas,  and S. Murciano for helpful discussions and collaboration on related topics. CR and PC are supported by the ERC under Consolidator Grant number 771536 (NEMO). EV acknowledges support from the CNRS-IEA and LYSM. 
\end{acknowledgements}

\appendix

\section{Derivation of the occupation functions $\vartheta_m(\lambda)$}
\label{sec:appendix}

In this appendix we derive the occupation functions $\vartheta_m(\lambda)$ from the tilted ferromagnetic state, both in the gapped and gapless phases. 
Our derivation follows the Boundary Quantum Transfer Matrix approach (BQTM), which was originally introduced to compute a quantity called the Loschmidt echo \cite{pozsgay2013dynamical,piroli2017from} following a quench from integrable initial states. It works as follows. First, the time evolution is discretized by making use of the Trotter-Suzuki formula. From there, time-evolved physical quantities are recast in terms of an auxiliary transfer matrix, the BQTM, acting on $N$ sites in the space direction, where $N$ is the Trotter number. In the thermodynamic limit, physical quantities are determined by the leading eigenvalue of the BQTM, which can be expressed by integrability methods in terms of a set of auxiliary Bethe rapidities.
More specifically, as was observed in \cite{piroli2017from}, the post-quench occupation functions $\vartheta_m(\lambda)$ are recovered by taking a ``zero time limit'' of the BQTM. In this limit, the auxiliary rapidities collapse to simple analytical value, and the BQTM eigenvalue becomes formally independent of $N$. 
It follows from this discussion that we can construct the occupation functions from a BQTM with $N=0$ sites, which is simply a scalar. For generality we will discuss properties of the BQTM with $N$ sites and arbitrary inhomogeneities $\{\xi_i\}_{i=1,\ldots N}$ in the following, and will use the simplification $N=0$ when specified. 
The occupation functions $\vartheta_m(\lambda)$ were previously obtained in the gapped regime \cite{piroli2016exact} following another approach (see also \cite{piroli2017from}), however our result in the gapless regime is new.

\textbf{The BQTM:}
The Boundary Quantum Transfer Matrix associated with the tilted ferromagnetic state was introduced in \cite{piroli2017integrable}. The dependence in the initial state is encoded in a pair of reflection matrices 
\bea
K^{\pm}(u)&=&K(u\pm \eta/2,\xi_{\pm},\kappa_{\pm},\tau_{\pm})\,\label{aux_k_matrix}\\
K(u,\xi,\kappa,\tau)&=&
\left(\begin{array}{cc}
\sinh\left(\xi+u\right)&\kappa e^{\tau} \sinh\left(2u\right) \\
\kappa e^{-\tau}\sinh\left(2u\right)&\sinh(\xi-u)
\end{array}\right)\label{k_matrix_nondiag}\,, 
\eea
where the various parameters are related to the $\theta$ through : 
\be
\sinh\alpha_{\pm}\cosh\beta_{\pm}=\frac{\sinh\xi_{\pm}}{2\kappa_{\pm}}\,,\quad \cosh\alpha_{\pm}\sinh\beta_{\pm}=\frac{\cosh\xi_{\pm}}{2\kappa_{\pm}}\,.
\label{eq:parametrization}
\ee
with
\be
\alpha^{\mp}=\pm\eta/2\,, \quad 
\beta_{\pm}=i\frac{\pi}{2}\,, \quad 
\tau_\pm =\pm \left(i\frac{\pi}{2}+r\right)\,, \quad 
e^r=\tan\frac{\theta}{2}\,.\label{fpar4}
\ee
For $N=0$, the BQTM is just a scalar, and reads : 
\be 
T(u) = \mathrm{tr}(K^+(u) K^-(u))
\ee 

\textbf{Gapped regime: }
The BQTM $T(u)$ can be used to built an infinite hierarchy of transfer matrices, defined recursively as \cite{zhou1996504,piroli2017from}
\bea
T_0(u) &=& 1  \nonumber\\
T_1(u) &=& T(u)  \nonumber\\
T_j(u) &=& T_{j-1}\left(u-\frac{\eta}{2}\right)T_{1}\left(u+(j-1)\frac{\eta}{2}\right)  - f\left(u + (j-3)\frac{\eta}{2}\right) T_{j-2}(u-\eta) \,, \qquad j\geq 2 \,,
\label{Tsys1}
\eea
where the function $f$ is defined as in \cite{piroli2017from}.
From (\ref{Tsys1}) one can derive the relation
\be 
T_j\left( u+\frac{\eta}{2} \right)T_j\left( u-\frac{\eta}{2} \right) 
= 
T_{j+1}( u ) T_{j-1}( u )  + \Phi_j(u) \,, \qquad j\geq 1
\ee
where 
\be 
\Phi_j(u) = \prod_{k=1}^j f\left( u - (j+2-2k)\frac{\eta}{2} \right)    \,.
\ee 
This allows to further define the $Y$ operators, $Y_0 =0$ and, for $j \geq 1$, 
\be
Y_j(u)=\frac{T_{j-1}(u)T_{j+1}(u)}{\Phi_j(u)}
\,,
\label{y_function}
\ee
which obey the following set of relations, known as Y system
\be
Y_j\left(u+\frac{\eta}{2}\right)Y_j\left(u-\frac{\eta}{2}\right)=\left[1+Y_{j+1}\left(u\right)\right]\left[1+Y_{j-1}\left(u\right)\right]\,.
\label{y_system}
\ee

As was observed in \cite{piroli2017from}, in the gapped regime the occupation functions $\vartheta_m(\lambda)$ can be directly obtained from the $Y$ functions associated with the leading BQTM eigenvalue (which, as explained above, can be obtained by directly taking a BQTM with $N=0$ sites) : 
\be 
\eta_m(\lambda) = Y_m(i\lambda) \,,  \qquad m=1,2\ldots 
\ee 
where $\eta_m(\lambda)=(\rho^t_m(\lambda)-\rho_m(\lambda)) / \rho_m(\lambda) = \frac{1}{\vartheta_m(\lambda)}-1$ are the filling fractions for the various quasiparticle species.
Explicitly,
\be 
\vartheta_1(\lambda) = \frac{1}{1+\eta_1(\lambda)}=
\frac{f(i\lambda - \tfrac{\eta}{2})} {t_1(i\lambda + \tfrac{\eta}{2})t_1(i\lambda - \tfrac{\eta}{2})}\,, 
\label{eta1gapped}
\ee 
where 
\bea 
t_1(i\lambda) &=& -\cosh (\eta )+\cos (2 \lambda )+\frac{1}{4} \text{sech}^2\left(\frac{\eta
   }{2}\right) \cosh (2 r) (\cos (4 \lambda )-\cosh (2 \eta ))
   \\ 
 f(i\lambda - \tfrac{\eta}{2}) &=& \frac{ \sin ^4(\lambda )}{8\cosh^4\left(\frac{\eta }{2}\right)}
   \tanh \left(\frac{1}{2} (\eta -2 i \lambda )\right) \tanh
   \left(\frac{1}{2} (\eta +2 i \lambda )\right) (\cosh (4 \eta )-\cos (4
   \lambda )) \,,
\eea 
while the further functions  $\eta_m(\lambda)$ follow from the Y system \eqref{y_system}. 
We check that this result agrees with a previous derivation using a different approach in \cite{piroli2016exact}.

\textbf{Gapless regime: }
We now turn to the gapless regime, and more specifically to anisotropies of the form $\Delta= \cos\gamma$, where $\gamma= \frac{\pi}{p+1}$, $p$ integer. 
At such values, the T and Y systems described in the previous paragraph can be truncated to a finite number of equations, which reflects the closure of the TBA over a finite set of quasiparticle types. 
This structure was described in \cite{piroli2018nonanalytic} in the case of a quench from the Néel state, and takes the form of a linear relation between $T_{p+1}$, $T_{p-1}$, and the identity,
\be 
t_{p+1}(u) = \alpha(u) t_{p-1}(u) + \beta(u) \phi(u+i \tfrac{\pi}{2}) \,,
\label{Ttruncation}
\ee 
where the coefficients $\alpha(u)$ and $\beta(u)$ will be specified below, and where we have 
introduced the rescaled transfer matrices
\be 
t_j(u) = \frac{T_j(u)}{\prod_{l=1}^{j-1} \phi(u+(2l-j)\eta)}\,, \qquad j\geq 1  \,. 
\label{tildetj}
\ee 
Here $\eta= i \gamma$ is now imaginary, and the function $\phi(u)$ contains all the explicit dependence in the number of sites $N$ and inhomogeneity parameters $\xi_i$ of the BQTM : it is defined as \cite{piroli2017from}
\be 
\phi(u) = \prod_{k=1}^{N}\sinh\left(u-\eta/2+\xi_k\right)\sinh\left(u+\eta/2-\xi_k\right)\,,\label{phi_function}
\ee 
and in particular is equal to $1$ for the case of a zero site BQTM. 

The structure \eqref{Ttruncation} is the same as that described in \cite{piroli2018nonanalytic}, however a novelty here is that the reflection matrices $K^\pm(u)$ associated with the tilted ferromagnet are non-diagonal, and as a result the coefficients $\alpha(u)$ and $\beta(u)$ are different and need to be computed explicitly.
In terms of the functions $\Phi_j$ and $f$ defined above, as well as the functions 
\bea 
\Psi(u) &=&  \frac{\sinh^2\left((p+1)(u+i\tfrac{\pi}{2})\right)\sinh^2\left((p+1)u\right)}{2^{4p-2}(\cosh\tfrac{\eta}{2})^{2(p-1)}}
\frac{\tanh(u+\tfrac{\eta}{2})\tanh(u-\tfrac{\eta}{2})}{\sinh(2u+2\eta)\sinh(2u-2\eta) \cosh^4\eta} \,.
\label{defPsi}
\\
\Phi(u) &=& \Psi(u) 
\times \phi\left( u-i\frac{\pi}{2} \right) \prod_{j=2}^{p-1} \phi\left( u+j\eta-i\frac{\pi}{2} \right) 
\eea 
we check : 
\bea 
\alpha(u) 
&=& \frac{\Phi_p(u)}{\Phi(u)^2} = \frac{\Phi(u-\eta)}{\Phi(u)} f(u+(p-2)\eta) 
\label{identityalpha}
\\ 
\beta(u) &=& - 2 \cosh(2(p+1)r) \Psi(u-\eta) f(u+(p-2)\eta)  \,.
\label{identitybeta}
\eea

We now turn to the truncation of the Y system. Using the following identity,
\be 
\Phi_{p-1}(u) = \Phi(u-\tfrac{\eta}{2})\Phi(u+\tfrac{\eta}{2}) \,,
\ee 
a first relation can be written under the form
\be 
1+ Y_{p-1}(u) = \frac{T_{p-2}(u)T_{p-1}(u) + \Phi_{p-1}(u)}{\Phi_{p-1}(u)}  
= \frac{T_{p-1}(u-\tfrac{\eta}{2})T_{p-1}(u+\tfrac{\eta}{2})}{\Phi(u+\tfrac{\eta}{2}) \Phi(u-\tfrac{\eta}{2})} 
= \mathcal{K}(u+\tfrac{\eta}{2}) \mathcal{K}(u-\tfrac{\eta}{2}) \,,
\ee 
where we have defined 
\be 
\mathcal{K}(u) =(-1)^{p+1} \frac{T_{p-1}(u)}{\Phi(u)} \,.
\label{Kdef}
\ee 

The last relation is obtained by plugging the truncation \eqref{Ttruncation} of the T system into the Y system relation 
\bea 
1+ Y_p(u) &=& 1+ \frac{T_{p-1}(u)T_{p+1}(u)}{\Psi_{p}(u)}  \,.
\eea 
Using the explicit expressions of $\alpha$ and $\beta$, we find
\be 
1 + Y_{p}(u) =  1 + 2 \cosh(2(p+1)r) \mathcal{K}(u) +  \mathcal{K}(u)^2  \,.
\ee 

We now specify to the leading eigenvalue of the BQTM, or alternatively consider a BQTM with $N=0$ sites, for which the $Y_j$ and $\mathcal{K}$ are now scalar functions. 
Following usual techniques, the truncated Y system can be recast in terms of a set of non linear integral relations for the functions $\{Y_j\}_{j=1,\ldots p}$ and $\mathcal{K}$ \cite{piroli2018nonanalytic}. Comparing these equations with the TBA relation obeyed by the filling fractions $\eta_m(\lambda)$ ($m=1,\ldots p+1$) \cite{takahashi1999thermodynamics} leads to the following identification 
\bea 
\eta_j(\lambda) &=& Y_j(\lambda)  \,, \qquad j=1, \ldots p-1 \\ 
\eta_{p}(\lambda) &=& e^{2(p+1)r} \mathcal{K}(\lambda)  \\
\eta_{p+1}(\lambda) &=& e^{2(p+1)r} / \mathcal{K}(\lambda)  \,.
\eea  

For concreteness, we give below the explicit form of the filling fractions for $p=1$ and $p=2$. 
For $p=1$, $\mathcal{K}(\lambda) =  \coth^2 \lambda$, and therefore  
\bea 
\eta_1(\lambda) &=& e^{4 r}\coth^2\lambda  
= \tan^4\frac{\theta}{2} \coth^2\lambda 
\\ 
\eta_2(\lambda) &=& e^{4 r}\tanh^2\lambda
= \tan^4\frac{\theta}{2} \tanh^2\lambda 
\,. 
\eea
From there, we obtain Eqs. \eqref{thetai_XX} in the main text.
For $p=2$, that is $\eta = \frac{i\pi}{3}$, we have
\bea 
\eta_1(\lambda) &=&  \mathcal{K}(\lambda +\tfrac{\eta}{2})\mathcal{K}(\lambda -\tfrac{\eta}{2}) -1 \\
\eta_2(\lambda) &=& e^{6 r} \mathcal{K}(\lambda) \\
\eta_3(\lambda) &=& e^{6 r} / \mathcal{K}(\lambda)  \,,
\eea 
with
\be 
\mathcal{K}(u) = \frac{2 \coth^2(u) (\cosh (2 r) (2 \cosh (2 u)+1)+3)}{2
   \cosh (2 u)-1} \,.
\ee

\bibliography{XXZbib.bib}

\end{document}